\documentclass[11pt]{article}
\usepackage[makeroom]{cancel}
\usepackage{cite}
\usepackage{graphicx}
\usepackage{multirow}
\usepackage{feynmf}
\usepackage[utf8]{inputenc}
\usepackage[english]{babel}
\usepackage{graphicx}
\usepackage{amsmath}
\usepackage{multicol,lipsum,graphicx,float}
\usepackage{color}
\usepackage{caption}
\usepackage{cite}
\usepackage{blindtext}
\usepackage{hyperref}
\usepackage{bbold}
\usepackage[bottom]{footmisc}
\makeatletter
\newcommand*{\rom}[1]{\expandafter\@slowromancap\romannumeral #1@}
\makeatother
\begin{document}
\begin{center}
{\Large\bf A Dynamic Modification to Sneutrino Chaotic Inflation}
%\\[.2mm]
%\vskip .2cm
\\
\vskip .5cm
{ Abhijit Kumar Saha$^{a,}$\footnote{abhijit.saha@iitg.ernet.in},
Arunansu Sil$^{a,}$\footnote{asil@iitg.ernet.in}}\\[3mm]
{\it{
$^a$ Indian Institute of Technology Guwahati, 781039 Assam, India}
}
\end{center}

\vskip .5cm

\begin{abstract}
We consider a right-handed scalar neutrino as the inflaton which carries a gravitational 
coupling with a supersymmetric QCD sector responsible for breaking supersymmetry dynamically. 
The framework suggests an inflaton potential which is a deformed version of the quadratic 
chaotic inflation leading to a flatter potential. We find that this deformation results a 
sizable tensor to scalar ratio which falls within the allowed region by PLANCK 2015. At the 
same time supersymmetry breaking at the end of inflation can naturally be induced in this 
set-up. The symmetries required to construct the framework allows the neutrino 
masses and mixing to be of right order.
\noindent \hspace{4cm}
\end{abstract}
%-------------------------------------------------------------------------

%\begin{multicols}{2}
\section{Introduction}
\label{sec:intro}

The inflationary paradigm is well accepted as a successful theory of the early universe
with its interpretation of several shortcomings of the Bigbang cosmology in an 
economic way. This hypothesis is further strengthened by its prediction
on the primordial perturbations that leads to the striking agreement with the observation
of the cosmic microwave background (CMB) spectrum. Among the various models of inflation,
large field inflation models receive a lot of attention these days, particularly after
the claim of BICEP2 \cite{Ade:2014xna}, due to their ability to produce a large tensor to scalar 
ratio ($r$). Although this particular claim is shadowed by the recent release of PLANCK 2015
\cite{Ade:2015xua,Ade:2015lrj} data which provides an upper bound on $r$ as $r \simeq0.11$,
large field inflation remains as an interesting possibility to explore in view of future search
for observing tensor perturbations in CMB by PLANCK 2015 and other experiments with a greater accuracy.

The chaotic inflation in supergravity (SUGRA) as proposed in \cite{Kawasaki:2000yn,Kawasaki:2000ws} having a scalar
 potential of the form $V=\frac{1}{2}m^2\chi^2$ is possibly the simplest scenario of this sort of inflation
 model\cite{Linde:2007fr} with its prediction of $r$ as $0.13$. The mass scale $m$ turns out to be of order $10^{13}$ GeV. Within this
large-field inflationary scenario, the inclusion of
supergravity induces a particular problem (known as the $\eta$ problem). This is caused by the field value
of the inflaton ($\chi$) during inflation, which exceeds the reduced Planck scale $M_P \simeq 2.4 \times
10^{18}$ GeV, and thereby spoiling the required degree of flatness of the inflationary potential
through the Planck-suppressed operators. In \cite{Kawasaki:2000yn}, a shift symmetric K$\ddot{\textrm{a}}$hler
potential associated with the inflaton field was considered to cure this problem. However the
PLANCK 2015\cite{Ade:2015xua,Ade:2015lrj} suggests a modification of the standard chaotic inflation as it barely enters into
the 2$\sigma$ range of $n_s - r$ (spectral index vs. tensor to scalar ratio) plot. Analyses
\cite{Harigaya:2014fca,Li:2013nfa} with PLANCK 2013\cite{Ade:2013zuv} data followed by the
BICEP2\cite{Ade:2014xna} suggest modification of the K$\ddot{\textrm{a}}$hler potential by introducing a shift-symmetry breaking
term. A general deformation of the chaotic superpotential by including higher order terms with
very small coefficients has been exercised in \cite{Nakayama:2013txa,Nakayama:2014wpa}. All these analyses would be further
restricted by the recent release of PLANCK 2015\cite{Ade:2015xua,Ade:2015lrj} data.

It has long been exercised how an inflationary scenario can be linked with the particle physics
framework. In this regard, neutrino physics can provide an interesting possibility. It is well
known that the smallness of the light neutrino mass ($m_{\nu}$) can be explained by type-I
seesaw mechanism which enforces the inclusion of heavy right handed (RH) neutrinos ($N$). In a supersymmetric theory, the close
proximity of the mass scale of these heavy RH fields and their superpartners (sneutrinos) with the mass parameter $m$ involved 
in the chaotic inflationary potential as mentioned before, suggests that the sneutrino can actually play the role for the inflaton.
 Indeed, it was shown \cite{Murayama:1992ua,Khalil:2011kd,Ellis:2004hy,Murayama:2014saa,Evans:2015mta} that the standard
chaotic inflation can actually be realized including them. Another interesting aspect of a
supersymmetric model of inflation is its relation with supersymmetry breaking. From the
completeness point of view, a supersymmetric structure of an inflationary scenario demands a
realization of supersymmetry breaking at the end of inflation. Though during inflation, the
vacuum energy responsible for inflation breaks supersymmetry (at a large scale of order of
 energy scale of inflation), as the inflaton field finally rolls down to a
global supersymmetric minimum, it reduces to zero vacuum energy, and thereby no residual
supersymmetry breaking remains.

In this work, our purpose is two fold; one is to modify the standard sneutrino chaotic inflation
so as to satisfy the PLANCK 2015\cite{Ade:2015xua,Ade:2015lrj} results and other is to accommodate supersymmetry breaking at the end of 
inflation. We consider two sectors namely (i) the inflation sector and (ii) the supersymmetry breaking sector. The inflation sector
is part of the neutrino sector consisting of three RH neutrino superfields. There will be a role for another sneutrino during 
inflation, which will be unfolded as we proceed. We identify the scalar field responsible for inflation to be associated with 
one of these three fields. The scalar potential resembles the standard
chaotic inflation in the supergravity framework assisted with the shift symmetric K$\ddot{\textrm{a}}$hler potential. The 
superpotential involving the RH neutrino responsible for inflation breaks this shift symmetry softly. We have argued the smallness 
associated with this shift symmetry breaking parameter by introducing a spurion field. In this excercise, we also consider discrete 
symmetries to forbid unwanted terms. For the supersymmetry breaking sector, we consider 
the Intriligator-Seiberg-Shih (ISS) model \cite{Intriligator:2006dd} of breaking supersymmetry dynamically in a metastable 
vacuum. This sector is described by a supersymmetric gauge theory and henceforth called the SQCD sector. These two 
sectors can have a gravitational coupling which in turn provides a dynamical deformation of the standard chaotic 
inflation. Again the coupling strength between these two sectors can be naturally obtained through another spurion. As the 
inflaton field approaches its minimum once the inflation is over, this interaction term becomes insignificant and finally 
the two sectors are effectively decoupled. However the hidden SQCD sector fields stabilize in metastable vacuum, hence supersymmetry 
breaking is achieved as a remnant of inflation.
Earlier attempts in connecting the inflation and ISS type supersymmetry breaking
can be found in \cite{Brax:2008rk, Savoy:2007jb, Brax:2009yd, Craig:2008tv}. A global $U(1)_R$ symmetry plays a pivotal role in shaping the 
ISS model of dynamic supersymmetry breaking. Once the supersymmetry is broken in the hidden SQCD sector, the effective supersymmetry breaking 
scale in the Standard Model sector  is assumed here to be developed by the gauge mediation mechanism. To materialize this, the ISS model requires
 a modification for breaking $U(1)_R$. In
this context we follow the proposal in \cite{Abel:2007jx} and show that this can easily be adopted in our 
set-up. Furthermore, as the RH neutrino superfields are part of the inflation sector, which obeys the same $U(1)_R$ symmetry, their $U(1)_R$ charges 
are already fixed. The same RH neutrinos also contribute to the light neutrino mass matrix through type-I seesaw mechanism. We find that the $U(1)_R$ 
charges of various fields involved along with their charges under the discrete symmetries imposed can actually predict an inverted hierarchy of 
neutrino masses. We provide an estimate of reheating temperature in this context and also comment on leptogenesis.

Below in section \ref{sec:2}, we briefly discuss the standard chaotic inflation in the supergravity framework.
Then we will discuss about the ISS model of dynamic supersymmetry breaking in section \ref{sec:3} followed by the role of
interaction term between the two sectors in section \ref{sec:4}. The dynamics of the fields during and after inflation are 
discussed in section \ref{sec:dyn1} and \ref{sec:dyn2} respectively. The prediction for this modified chaotic inflation
 are presented in section \ref{sec:6}. In section \ref{sec:R break}, we have shown that a deformation to the SQCD sector
 can be achieved which is related to the $U(1)_R$ symmetry breaking. In section \ref{sec:9}, we discuss the implication
 of neutrino masses and mixing that comes out of the present set-up. We comment on the reheating temperature and
 leptogenesis in section \ref{sec:10} . Finally we conclude in section \ref{sec:conclude}.
%--------------------------------------------------------------------------Inflation Sector--------------------------------------------------------------------------------
%--------------------------------------------------------------------------------------------------------------------------------------------------------------------------
\section{Standard sneutrino chaotic inflation in supergravity}\label{sec:2}
We start this section by reviewing some of the features of the standard chaotic inflation
in supergravity where the scalar partner of a RH neutrino (say $N_1$ among the three RH
superfields $N_{i=1,2,3}$ involved in type-I seesaw for generating light neutrino mass) serves the role of inflaton. Sneutrino chaotic
inflation\cite{Murayama:1992ua, Murayama:2014saa,Ellis:2004hy,Evans:2015mta} gains much attention from the perspective
of particle physics involvement. Mass of the inflaton and in turn mass of that particular
RH neutrino (in the supersymmetric limit) can be fixed by the magnitude of
curvature perturbation spectrum in this theory. In $\mathcal{N}=1$ SUGRA, the superpotential
is considered to be
\begin{equation}\label{eq:sup1}
W_N=m N_1 N_2,
\end{equation}
along with the K$\ddot{\textrm{a}}$hler Potential\footnote{$K_N$ also involves $|N_3|^2$, which we do not put here for simplifying our discussion. }
\begin{equation}\label{eq:Kahler1}
K_N=|N_2|^{2}-\frac{(N_1-N_1^{\dagger})^{2}}{2}.
\end{equation}
Note that a shift symmetry, $N_1\rightarrow N_1+\textrm{C}$, where C is  real having mass dimension unity, is imposed on the K$\ddot{\textrm{a}}$hler potential,
whereas the superpotential breaks it. Thus the parameter $m$ can be regarded
as a shift-symmetry breaking parameter.

The parameter $m$ being much smaller than $M_P$, the term in 
the superpotential $W_N$ would be natural in $^{,}$t Hooft's sense \cite{Hooft:1980} of increased 
symmetry in the limit $m \rightarrow$ 0. The smallness associated with $m$ can 
be explained with the introduction of a spurion field $z_1$ as shown in \cite{Kawasaki:2000ws}. 
Also the higher order shift symmetry breaking terms involving $N_1$ can be controlled in an 
elegant way through the introduction of $z_1$. Suppose the spurion field $z_1$ transforms 
under the shift symmetry as, 
\begin{equation}
z_1 \rightarrow \frac{N_1}{N_1+C} z_1,
\end{equation}    
hence $N_1 z_1$ combination remains shift symmetric. 
At this stage, a discussion on $U(1)_R$ symmetry is
pertinent. There exists a global $U(1)_R$ symmetry under which the superpotential
$W$ has 2 units of $R$-charges. However note that with the presence of shift symmetric
K$\ddot{\textrm{a}}$hler potential involving $N_1$, $N_1$ can not possess a
global $U(1)_R$ charge. Therefore $N_2$ should carry $R$-charge 2, while
$R$-charges of $N_1$ and $z_1$ are zero. Furthermore, we consider a $\textrm{Z}_2$ symmetry 
under which only $N_1$ and $N_2$ are odd. Combining the shift symmetry, $U(1)_R$ and the $\textrm{Z}_2$ (charges are specified in Table \ref{tab:lepton}), we can write the general 
superpotential for $W_N$ as 
\begin{equation}
W^g_N = [z_1 N_1 + a_3 (z_1 N_1)^3 + ... ] N_2.
\end{equation}
As the $z_1$ gets a vacuum expectation value (vev) $\sim m$ which is small compared to $M_P$, we can 
argue that the shift symmetry is softly broken. Simultaneously the higher order 
terms (with coefficient $a_{i}\sim\mathcal{O}(1)$) are negligibly small and hence we are essentially left with 
our working superpotential $W_N$ in Eq.(\ref{eq:sup1}).

The importance of having this shift symmetry
can be understood as discussed below. F-term scalar potential is calculated using the
following standard expression,
\begin{equation}\label{eq:sug1}
V_{F}=e^{\frac{K}{M_{P}^2}}\Big[D_i WK_{ij^*}^{-1}D_{j^{*}}W^*-3\frac{|W|^2}{M_{P}^2}\Big],
\end{equation}
where $D_i W=\frac{\partial W}{\partial f_i}+K_i/M_{P}^2$ and the subscript
$i$ labels a superfield $f_i$. Due to the imposed shift symmetry on $N_1$, the K$\ddot{\textrm{a}}$hler
potential (or $e^{K/M_{P}^2}$) depends only on the imaginary component
of $N_1$. The real component of $N_1$ therefore can be considered to be the inflaton (hereafter denoted by $\chi$). Its absence in the K$\ddot{\textrm{a}}$hler potential allows it
to acquire super-Planckian value during inflation, which is a characteristic of large field inflation models. Assuming that during inflation, all other fields (including $N_2$ as well)
except the inflaton are stabilized at origin\footnote{Particularly for $N_2$,
this can be ensured by adding a non-canonical term in the K$\ddot{\textrm{a}}$hler as
$\xi |N_2|^{4}/\left[ 2 M^2_P\right]$ with $\xi \sim 1$ \cite{Kawasaki:2000yn}.}, the inflationary
potential becomes $V_{\chi}=\frac{1}{2}m^2\chi^2$. The standard slow roll parameters are
defined as
\begin{equation}
\epsilon=\frac{M_P^2}{2}\Big(\frac{V^{\prime}}{V} \Big)^2  \textrm{ ; } \eta=\frac{M_P^2 V^{\prime\prime}}{V},
\end{equation}
where $\prime$ denotes the derivative of the potential with respect to inflaton field.
Number of $e$-foldings can be calculated by the following,
\begin{equation}
N_e=\frac{1}{M_P^2}\int_{\chi_{end}}^{\chi_{*}}\frac{V}{V^{\prime}} d\chi.
\end{equation}
Other cosmological observables like spectral index ($n_s$), tensor to scalar ratio ($r$), curvature perturbation spectrum($P_{\zeta}$) are given by
\begin{equation}
n_s=1-6\epsilon+2\eta; \textrm{ } r=16\epsilon; \textrm{ } P_{\zeta}=\frac{V}{24 M_P^4 \pi^2 \epsilon},
\end{equation}
respectively. Chaotic inflation with $V_{\chi}$ then predicts
\begin{equation}
n_{s}\simeq0.967  \textrm{        and    }   r\simeq0.13 ,
\end{equation}
where $N_{e}=60$ is considered. Inflation starts at $\chi^{*}=15.5 M_P$ and ends at $\chi^{end}=\sqrt{2}M_P$. The value of $m$ turns out
to be $\sim 10^{13}$ GeV so as to produce the correct order of curvature perturbation spectrum
$P_{\zeta}=2.2\times10^{-9}$. Note that this $m$ falls in the right ballpark for generating light neutrino mass through type-I seesaw. However in view of the recent PLANCK update \cite{Ade:2015xua,Ade:2015lrj}, this minimal
model is almost outside  the 2$\sigma$ region of $n_s-r$ plot. So a modification of the
minimal model is of utmost importance. As we have mentioned before, there has been
some suggestions toward this \cite{Harigaya:2015pea,Pallis:2015mga, Barenboim:2015lla,Carpenter:2014saa,Heurtier:2015ima}. In this work, our approach to accommodate chaotic
inflation within the present experimental limit is to couple it with the supersymmetry
breaking sector. This coupling serves as a dynamic modification to the minimal chaotic inflation.
To discuss it in detail, in the following section we present a brief summary of the ISS
model of dynamical supersymmetry breaking.

%----------------------------------------------------------ISS Sector-----------------------------------------------------------
%--------------------------------------------------------------------------------------------------------------------------------
\section{SQCD sector and supersymmetry breaking in a metastable vacuum }\label{sec:3}

It is evident from the F-terms (in particular $ F_{N_2} = m N_1$) of $W_N$ in
Eq.(\ref{eq:sup1}) that during inflation, supersymmetry is broken at a very
high scale since the inflaton ($\chi$ field $\equiv$ real part of $N_1$)
takes a non-zero super-Planckian value. However once the inflation is
over, the $\chi$ field finally acquires a field-value zero ($\chi=0$ is the global minimum) as evident from the
minimization of the potential
$V_{\chi} = m^2 \chi^2/2$. Hence there is no supersymmetry breaking associated with this minimum. It is expected that there should be a small amount of supersymmetry breaking left at the end of
inflation so that an effective supersymmetry breaking in the supersymmetric version of the Standard model or its extension can be introduced. In this work, we consider the inflation sector to be assisted by a separate hidden sector
responsible for supersymmetry breaking\footnote{Another 
approach to accommodate supersymmetry breaking after chaotic inflation is exercised in \cite{Buchmuller:2014pla} with an introduction of 
a Polonyi field.}. We consider the hidden sector to be described by a supersymmetric gauge theory similar to the one considered in the ISS model of dynamic supersymmetry breaking \cite{Intriligator:2007cp}. Recently a
proposal \cite{Harigaya:2014wta,Harigaya:2014sua} of generating chaotic potential for a strongly interacting supersymmetry gauge theory is analysed which leads to a fractional chaotic inflation. However in our approach we consider the SQCD sector 
to provide a deformation to the sneutrino contribution to the minimal chaotic inflation, and at the end of inflation, this serves as the hidden sector of the supersymmetry breaking. The effective
supersymmetric breaking in the standard supersymmetric gauge and
matter sector (MSSM or its extension) requires a mediation mechanism from this hidden sector. Here it is considered to be the gauge mediation.

The ISS model is described by the $\mathcal{N}=1$ supersymmetric $SU(N_{C})$ gauge theory
(called the electric theory) with $N_{f}$ flavors of quarks ($Q$) and
antiquarks $(\tilde{Q})$. $\Lambda$ is the strong coupling scale of this theory. Below
this scale $\Lambda$, the theory is described by its magnetic dual
$SU(N= N_{f}-N_{C})$ gauge theory with $N_f$ flavors of magnetic quarks $q_i^c,
\tilde q^c_i$ (with
$i=1....N_{f}$ and $c=1....N$). It is interesting to note that this theory is IR free, provided $N_{C}+1\leq N_{f}<
\frac{3}{2}N_{C}$. The elegance of the ISS model lies in its UV completion of the theory.
There also exists a $N_f \times N_f$ gauge singlet meson field $\Phi = Q
\tilde Q/\Lambda$. With the introduction of quark mass term in the electric theory ($SU(N_c)$ gauge theory),
\begin{equation}\label{eq:quark}
W_e = m_Q{\rm{Tr}}Q \tilde Q,
\end{equation}
with $m_Q<\Lambda$, the IR free magnetic theory becomes
\begin{equation}\label{eq:ISS1}
W_{ISS}=h{\rm{Tr}}(q\Phi\tilde{q})-h\mu^{2}{\rm{Tr}}(\Phi),
\end{equation}
along with the dynamical superpotential
\begin{equation}\label{eq:dyn}
W_{dyn}=N\Big(h^{N_{f}}\frac{\textrm{det}\Phi}{\Lambda^{N_{f}-3N}}\Big)^{\frac{1}{N}}.
\end{equation}
where $h \sim \mathcal{O}(1)$ and $\mu \ll  \Lambda$ and by duality $\mu^2=m_{Q}\Lambda$.  Note that there exists a 
$U(1)_R$ symmetry under which $W_{ISS}$ and hence $\Phi$ carry $R$-charge of 2-units. R charge of $Q \tilde{Q}$ 
combination turns out to be two as well from the relation $\Phi = Q\tilde Q/\Lambda$. However the $R$ symmetry 
is explicitly broken by the  $W_{dyn}$  term. All the fields in this sector are considered to be even under the $\textrm{Z}_2$ symmetry considered. The K$\ddot{\textrm{a}}$hler potential is considered to be canonical in both electric and magnetic theories. It is shown in \cite{Intriligator:2006dd} that there exists a local minimum given by
\begin{equation}\label{eq:meta}
\langle q \rangle=\langle \tilde{q}^{T} \rangle=\mu  \left( \begin{array}{c}
\mathbb{1}_{N} \\
0_{N_{f}-N}   \end{array} \right) ,\textrm{  } \langle \Phi \rangle=0,
\end{equation}
with vacuum energy $V_{ISS}=N_c|h^{2}\mu^{4}|$. Supersymmetry is broken in this minimum by the rank condition. Note that $W_{dyn}$ is almost negligible around $\Phi=0$. The interplay between second term in Eq.(\ref{eq:ISS1}) and the $W_{dyn}$ suggests an existence of a SUSY preserving vacuum at
\begin{equation}
\langle q\rangle=\langle \tilde{q}^{T} \rangle=0, \textrm{  } \langle\Phi\rangle=\frac{\mu}{h}(\epsilon^{\frac{N_{f}-3 N}{N_{c}}} )^{-1}\mathbb{1}_{N_{f}},
\end{equation}
where $\epsilon=\frac{\mu}{\Lambda}$ and the corresponding vacuum energy $V_0=0$. With $\epsilon\ll1$, it was shown in \cite{Intriligator:2006dd} that the local minima in Eq.(\ref{eq:meta}) is a metastable one.

%------------------------------------Interaction between two sector------------------------------------------------------------
%-------------------------------------------------------------------------------------------------------------------------------
\section{Interaction between neutrino and SQCD sectors}\label{sec:4}

We consider $W_{N}$ as the superpotential describing the inflation with $N_1$
playing the role of inflaton. In this section our endeavor is to couple the inflaton with the SQCD sector. We
assume that the two sectors can communicate with each other
only through gravity. The lowest dimensional operator consistent with the set-up is therefore given by,
\begin{equation}
W_{\rm{Int}}=\beta \frac{N_1^{2}{\rm{Tr}}(Q\tilde{Q})}{M_{P}},
\label{wint}
\end{equation}
where $\beta$ is a coupling constant. We consider $\beta$ to be much less than unity. Similar to $W_{ISS}$, 
$W_{\rm{Int}}$ also respects the $U(1)_R$ and hence linear in Tr$(Q\tilde{Q})$ having $R$-charge 2. Among $N_1$ and $N_2$, it is therefore the $N_1$ field 
only which can couple ($N_2$ carries
2 units of $R$-charge) with the ISS sector. Since 
the interaction between the two sectors are assumed to be mediated by gravity only, the interaction term is expected to be $M_P$ suppressed. 
Hence $W_{\textrm{Int}}$ in Eq.(\ref{wint}) serves as the minimal description of the interaction between the two sectors. Being a shift-symmetry
 breaking parameter, the origin of $\beta$ can be 
explained with the introduction of another spurion field $z_2$ which transforms as 
$z_2 \rightarrow \frac{N^2_1}{(N_1  +  C)^2} z_2$ under shift symmetry. We consider $z_2$ to be even under the $\textrm{Z}_2$ symmetry and it does not 
carry any $R$ charge. On the other hand, $Q\tilde{Q}$ combination is even under $\textrm{Z}_2$. We introduce another discrete 
symmetry $\textrm{Z}_4$ under which $z_2$ carries a charge $i$ as well as $Q\tilde{Q}$ carrying charge $-i$. Hence $m_Q$ also 
carries a $\textrm{Z}_4$ charge $i$ as seen from Eq.(\ref{eq:quark}). Application of this symmetry forbids dangerous term 
like $\frac{(z_2N_1^2)z_1N_1N_2}{M_P^3}$. Hence a general superpotential involving $z_2 N_1^2$ can be obtained as 
\begin{align}
W_{\textrm{Int}}^{g}=\frac{\textrm{Tr}(Q\tilde{Q})}{M_{P}} \Big[\frac{z_2 N^2_1}{M_P}  + b_5 \frac{(z_2 N^2_1)^5}{M_P^12} +.....\Big],
\end{align}
where $b_{5}$ corresponds to respective coupling. Terms involving quadratic, cubic, and quartic powers of $(z_2 N_1^2)$ are not 
allowed from the $\textrm{Z}_4$ charge assignment as considered\footnote{With this new $\textrm{Z}_4$ symmetry, term 
like $\frac{(z_2 N_1^2)z_1N_1N_2}{M_P^3}$ will be allowed in $W_{\textrm{Int}}$. However contribution of this term will 
be negligibly small.}. Therefore $\beta$ is obtained through $\beta=\langle z_2 \rangle/M_P$. Note that 
with $\beta\sim 10^{-3}$ (as we will see), terms with $b_5$ and higher orders are negligibly small.

%Hence a general superpotential involving $z_2 N^2_1$ can be obtained as, 
%\begin{equation}
%W_{\rm{Int}}  = \frac{\textrm{Tr}(Q\tilde{Q})}{M_{P}} \Big[\frac{z_2 N^2_1}{M_P}  + b_2 \Big(\frac{z_2 N^2_1}{M_P}\Big)^2  ...\Big],
%\end{equation}
%where $b_{i_{=2,4,..}}$ correspond to respective couplings. 
%The coefficient $\beta$ in Eq. (\ref{wint}) can be obtained when $z_2$ gets a VEV so that  $\beta =  \frac{\langle z_2 \rangle}{M_P}$. Once $\beta$ is considered to be sufficiently smaller than one, higher 
%order terms are negligible for our purpose. Note that the introduction of $z_2$ spurion does not introduce any 
%new contribution to the $W_N$.

Note that this interaction term in addition to the quark mass term $m_Q$ present in $W_e$ (see Eq.(\ref{eq:quark})),
generates an effective mass for the electric quarks, $m'_Q=\beta \frac{N_1^{2}}{M_{P}}
+ m_Q$. Here we are particularly interested in the case when the
effective mass of the quarks, $m'_Q$, becomes larger than the cut-off
scale $\Lambda$, {\it {i.e.}} when $ m'_Q \gg \Lambda$.  Since $m_Q$ is considered to be less than $\Lambda$ in the ISS set-up, 
this situation can be achieved when the inflaton field $N_1$ satisfies, $N_1 \gg \left [ \Lambda M_P / \beta \right]^{1/2}$.
%\cite{Intriligator:2007cp},
These heavy quarks can then be integrated out \cite{Intriligator:2007cp}
to form an effective theory with a field dependent dynamical scale, $\Lambda_{\rm{eff}}(N_1)$. As all the 
quarks are getting large masses, the effective theory
becomes a pure gauge theory with no flavors. $\Lambda_{\rm{eff}}$,
can be determined by the standard scale matching of the gauge
couplings of two theories at an energy scale $E=m_Q^{\prime}$.  With $g_{N_c,N_f}$ and $g_{N_c,0}$ are the gauge couplings of the $SU(N_c)$ 
gauge theory with $N_f$ flavors of quarks ($Q,\tilde{Q}$) $(E>m_Q^{\prime})$ and pure gauge theory with $N_f=0$ $(E<m_Q^{\prime})$ respectively, the 
condition $g_{N_{c},N_{f}}(m_Q^{\prime})=g_{N_c,0}(m_Q^{\prime})$ gives
\begin{equation}
\Big(\frac{m_{Q}^{\prime}}{\Lambda}\Big)^b=\Big(\frac{m_{Q}^{\prime}}{\Lambda_{\rm{eff}}}\Big)^{b_{eff}},
\end{equation}
where $b=3N_c-N_f$ and $b_{eff}=3N_c$ are the respective beta functions of gauge couplings of the two theories. $\Lambda_{eff}$ in our set up turns out to be
\begin{equation}
\Lambda _{{\rm{eff}}}\simeq\Big(\frac{\beta N_1^{2}}{M_{P}}\Big)^{(1-p)}\Lambda^{p},
\end{equation}
where $p=\frac{b}{b_{\rm{eff}}}$ and $m_Q^{\prime}$ is mostly dominated by $\frac{\beta N_1^2}{M_P}$ term ({\it{i.e.}} 
when $N_1 \gg \left [ \Lambda M_P / \beta \right]^{1/2}$ and $m_{Q}$ being much smaller than $\Lambda$ can be neglected).  As 
all the flavors are integrated out, the superpotential describing the effective theory is generated
via gaugino condensation and is given by\cite{Intriligator:2007cp}
\begin{equation}
W_{\rm{Int}}^{\rm{eff}}= N_{c}\Lambda _{\rm{eff}}^{3}=N_{c}\Big(\beta\frac{N_1^{2}}{M_{P}}\Big)^{3(1-p)}\Lambda^{3p}.
\end{equation}
Below we study the impact of this term on inflation governed by $W_N$.
%\noindent [Write the origin of the matching condition as equation ]

%---------------------------------------------Modified Chaotic inflation---------------------------------------------------------------------------------------------------
%---------------------------------------------------------------------------------------------------------------------------------------------------------------------
\section{Modified chaotic potential and its implications to inflationary dynamics}\label{sec:dyn1}
Here we will study the inflationary dynamics based on the superpotential,
\begin{equation}\label{eq:effW}
W_{\textrm{Inf}} = W_N+W_{\rm{Int}}^{\rm{eff}}=  m N_1 N_2+N_{c}\Big(\frac{\beta N_1^{2}}{M_{P}}\Big)^{3(1-p)}\Lambda^{3p},
\end{equation}
when $N_1\gg \left [ \Lambda M_P / \beta \right]^{1/2}$.
Note that it indicates a modification of the chaotic inflationary potential $V_{\chi}$ obtained from $W_N$ only.
In this section, we will study the outcome of this modified superpotential in
terms of prediction of parameters involved in inflation. Depending upon $p$, the superpotential may contain fractional 
powers of $N_1$. In Ref\cite{Gao:2014fha}, superpotential with
non-integer power of superfields has been studied. It is shown there that the form of
K$\ddot{\textrm{a}}$hler potential remains same irrespective of integer or non-integer power of
superfields involved in the superpotential.

The K$\ddot{\textrm{a}}$hler potential is considered to be the same as $K_N$ in Eq.(\ref{eq:Kahler1}).
We can write
\begin{equation}
N_1=\frac{\chi+i \eta}{\sqrt{2}} \textrm{     and        } N_2=\frac{\sigma+i\delta}{\sqrt{2}}.
\end{equation}
As discussed in section \ref{sec:2}, we choose $\chi$, the real component of $N_1$, as inflaton.
%For the completness, we will also study the effect of introducing a shift-symmetry breaking term in the Kahler potential so as to read,
%\begin{equation}
%K_m = K + \alpha \frac{(N_1+N_1^{\dagger})^{2}}{2},
%\end{equation}
%where $\alpha$ is the shift symmetry breaking parameter and $|\alpha\sigma^{2}| \ll 1$.
Using Eq.(\ref{eq:Kahler1},\ref{eq:sug1}) and Eq.(\ref{eq:effW}), the scalar potential involving $\chi$ and $\sigma$ is given by
%\begingroup
   % \fontsize{8.2pt}{10pt}\selectfont
%\begin{equation}
%W_{Inf} \sim m^{\prime}Tr(Q\bar{Q})+N_{c}\Big(\frac{N_1^{2}}{M_{P}}\Big)^{3(1-p)}\tilde{\Lambda}^{3p}+ m N_1 N_2
%\end{equation}
%\begin{align}\label{eq:Potmain}
%V_{scalar} &=e^{\frac{\tilde{\sigma}^2}{2 M_P^2}}\Big(\frac{m^2\tilde{\sigma}^2}{2}(1-\alpha)+\nonumber\\
%& \frac{m^2\chi^2}{2}\Big\{1+\frac{2\tilde{\sigma}^2}{M_P^2}(\frac{3}{2}\alpha-\frac{1}{4})+\frac{\tilde{\sigma}^4}{4M_P^4} \Big\}+\nonumber\\
%& \frac{m^2\chi^4\alpha}{2 M_P^2}\Big\{ 1-\frac{\tilde{\sigma}^2}{2 M_P^2}+\frac{\tilde{\sigma}^4}{4 M_P^4}\Big\} +\nonumber\\
%& \chi^{5-6p}\Big\{\frac{3 N_c(1-p)\tilde{\Lambda}^{3p}m\tilde{\sigma}}{M_P^{3(1-p)}2^{1-3P}}(1-\alpha)\Big\} )\nonumber\\
%& \chi^{7-6p}\frac{m N_c \tilde{\Lambda}^{3p}}{M_P^{5-3p}2^{1-3p}}\Big\{ \tilde{\sigma}(\frac{13\alpha}{2}-6\alpha p-\frac{1}{2})+\frac{\tilde{\sigma}^3}{8M_P^2}\Big\}+\nonumber\\
%& \chi^{9-6p}\frac{\alpha m N_c \tilde{\Lambda}^{3p}}{M_P^{7-3p}2^{2-3p}}\Big\{-\tilde{\sigma}+\frac{\tilde{\sigma}^3}{4 M_P^2}\Big\}+\nonumber\\
%& \chi^{10-12p}\frac{9 N_c^2(1-p)^2\tilde{\Lambda}^{6p}}{M_P^{6(1-p)}2^{3-6p}}(1-\alpha)+\nonumber\\
%& \chi^{12-12p}\frac{N_c^2\tilde{\Lambda}^{6p}}{M_P^{8-6p}2^{3-6p}}\Big\{12\alpha-21\alpha p+9\alpha p^2-\frac{3}{8}+\frac{\tilde{\sigma}^2}{16 M_P^2}  \Big\}+\nonumber\\
%& \chi^{14-12p}\frac{\alpha N_c^2\tilde{\Lambda}^{6p}}{M_P^{10-6p}2^{6-6p}}\Big\{ -3+\frac{\tilde{\sigma}^2}{2 M_P^2}\Big\}\Big)
%\end{align}
%\endgroup
\begin{align}\label{eq:Pot2}
V_{\rm{Inf}}(\tilde{\chi},\tilde{\sigma}) &= M_P^4e^{\tilde{\sigma}^2/2}\Big[\frac{\tilde{m}^2\tilde{\sigma}^2}{2}+\frac{\tilde{m}^2\tilde{\chi}^2}{2}\Big(1-\frac{\tilde{\sigma}^2}{2}+\frac{\tilde{\sigma}^4}{4}\Big)\nonumber\\
& +3(1-p)A\tilde{\chi}^{5-6p}\tilde{\sigma}+A\tilde{\chi}^{7-6p}\Big(-\frac{\tilde{\sigma}}{2}+\frac{\tilde{\sigma}^3}{8}\Big)\nonumber\\
& +\frac{9}{2\tilde{m}^2}(1-p)^2A^2\tilde{\chi}^{10-12p}+\frac{A^2}{2\tilde{m}^2}\tilde{\chi}^{12-12p}\Big(-\frac{3}{8}+\frac{\tilde{\sigma}^2}{16}\Big)  \Big],
\end{align}
%\vspace{5mm}
where $A=\frac{\beta^{3(1-p)} \tilde{m} N_c\tilde{\Lambda}^{3p}}{2^{1-3p}}$ and tilde indicates that the 
corresponding variable or parameter is scaled in terms of $M_P$, {\it{e.g.}} $\tilde{\sigma}=\frac{\sigma}{M_P}$. We follow 
this notation throughout this section only. As the electric quarks degrees of freedom ($Q,\tilde{Q}$) are integrated out we 
will not consider quarks anymore, as long as $N_1\gg\sqrt{\frac{\Lambda M_P}{\beta}}$. Now one can wonder what happened to 
other two fields $\eta$ and $\delta$. Due to the presence of $e^{K}$ factor in the scalar potential $V_{F}$
the effective mass of $\eta$ during inflation will be large compared to inflaton mass($m_{\eta}^2\sim 6 H_{\rm{Inf}}^2+m^2$) 
and hence it will quickly settle down to origin. We have checked numerically that the other field $\delta$ also settles at 
origin during inflation, having mass more than the Hubble.
\\

\begin{figure}[h]
%\hspace{.5cm}
\centering
\includegraphics[width=7cm]{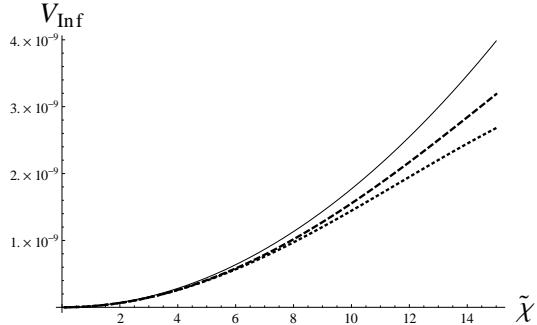}
\caption{Examples of inflation potential ($V_{\rm{Inf}}$) against $\tilde{\chi}$ are presented. The solid, large-dashed and small-dashed curves represent (I) minimal chaotic potential with $V_{\rm{Inf}}(\tilde{\chi})=\frac{1}{2}\tilde{m}^2\tilde{\chi}^2$, (II) modified $V_{\rm{Inf}}(\tilde{\chi})$ obtained from our set-up with $\alpha=0$ and (III) modified $V_{\rm{Inf}}(\tilde{\chi})$ with $\alpha=7\times 10^{-4}$ respectively. }
\label{fig:Pot1}
\end{figure}

In this case, dynamics of inflation belongs to $\tilde{\chi}-\tilde{\sigma}$ plane. Note that in case of standard chaotic inflation 
as discussed in section \ref{sec:2}, the $\tilde{\sigma}$ field is considered to be at origin during inflation. Contrary to that, 
the dynamic modification of the scalar potential governed by $W_{\rm{Int}}^{\rm{eff}}$ forces the $\tilde{\sigma}$ field to have a 
nonzero vacuum expectation value in our case. Similar type of scenarios are discussed in \cite{Evans:2015mta,Harigaya:2015pea}. In 
order to get $\langle\tilde{\sigma}\rangle$ in terms of $\tilde{\chi}$, derivative of the scalar potential with respect to $\tilde{\sigma}$ and yields
\begin{align}\label{eq:Potder}
\frac{\partial V_{\rm {Inf}}}{\partial \tilde{\sigma}} = e^{\frac {\tilde{\sigma}^2}{2}}\Big[-\frac{A\tilde{\chi}^{5-6p}}{2}(\tilde{\chi}^2-6+6p)+\tilde{\sigma}\Big\{ \tilde{m}^2+\tilde{\chi}^{10-12p}\frac{A^2}{2\tilde{m}^2}\Big(9-18p+9p^2-\frac{\tilde{\chi}^2}{4}\Big)\Big \}\nonumber \\+\tilde{\sigma}^2\Big \{A\tilde{\chi}^{5-6p}\Big(3-3p-\frac{\tilde{\chi}^2}{8}\Big)\Big\} +\tilde{\sigma}^3\Big\{\frac{\tilde{m}^2}{2}\Big(1+\frac{\tilde{\chi}^2}{2}\Big)+\frac{\tilde{\chi}^{12-12p}}{32}\frac{A^2}{\tilde{m}^2}\Big \}\nonumber \\+ \tilde{\sigma}^4 \Big \{ \tilde{\chi}^{7-6p}\frac{A}{8}\Big \} + \tilde{\sigma}^5 \Big \{ \frac{\tilde{\chi}^{2}\tilde{m}^2}{8}\Big \} \Big].
\end{align}
In order to minimize the scalar potential, we equate the above expression to zero ({\it{i.e.}} $\frac{\partial V_{\rm{Inf}}(\tilde{\chi},\tilde{\sigma})}{\partial \tilde{\sigma}}$=0). It 
reduces to a fifth order polynomial equation in $\tilde{\sigma}$. At this point we consider a specific value of $p \textrm{ }(=1-\frac{N_f}{3 N_c})$, 
the choice of which is guided by the construction of the ISS framework and a possible realization of $U(1)_R$ breaking through baryon deformation 
as we will discuss in section \ref{sec:R break}. Comparing the relative magnitudes of the terms involved in the fifth order polynomial and 
considering $\tilde{\sigma}$ to be sub-Planckian, we solve the equation for $\tilde{\sigma}$ in a perturbative way, the details of which 
is given in Appendix \ref{App:Appendix1}. Once $\langle\tilde{\sigma}\rangle$ is obtained in terms of $\tilde{\chi}$, we replace $\tilde{\sigma}$ 
by its VEV in Eq.(\ref{eq:Pot2}) and potential responsible for inflation now becomes function of $\tilde{\chi}$ only. Due to its very 
complicated functional dependence on $\tilde{\chi}$, we have not presented $V_{\textrm{Inf}}$ here. Instead in Fig.\ref{fig:Pot1} we have
depicted the potential $V(\tilde{\chi})$ in terms of $\tilde{\chi}$ for $p=4/7$ (indicated by dashed line). Note that this potential 
is indeed flatter compared to the standard sneutrino
chaotic inflation potential \cite{Murayama:2014saa}, indicated in Fig.\ref{fig:Pot1} by the solid line.
%$\frac{\partial V}{\partial\tilde{\sigma}}=0$ is a fifth order polynommial equation which is difficult to solve analytically. We use perturbation technique to solve this, the details of which can be found in Appendix.Then we replace the $\langle\tilde{\sigma}\rangle$ into the supergravity scalar potential$(V_{Inf}^{\tilde{\chi},\tilde{\sigma}})$ in Eq.\ref{eq:Pot2}  to get the inflationary potential$(V_{inf}(\tilde{\chi}))$. As seen from the Fig.\ref{fig:Pot1} that $V_{Inf(\tilde{\chi})})$(Brown line) is flatter\footnote{A potential $V_{1}$ is flatter than another potential $V_2$ means, for a particular value of $\tilde{\chi}$ field $V_2<V_1$} than the usual standard chaotic inflationary potential(Red line).

\hspace{1cm}

For completeness a shift symmetry breaking parameter ($\alpha$) in  the K$\ddot{\textrm{a}}$hler potential can also be introduced. The modified K$\ddot{\textrm{a}}$hler potential will look like as
\begin{equation}
K = K_N - \alpha \frac{(N_1+N_1^{\dagger})^{2}}{2},
\end{equation}
with $\alpha\ll1$. The scalar potential in Eq.(\ref{eq:Pot2}) will be modified and takes a further complicated form. In this case we obtain the scalar potential as a function of $\tilde{\chi}$ in a similar way. In Fig.(\ref{fig:Pot1}) we also plot $V_{\rm{Inf}}(\tilde{\chi})$ including the nonzero value of $\alpha\{\sim7\times10^{-4}\}$ represented by dashed line. It is to be noted that introduction of $\alpha$ makes the shape of $V_{\rm{Inf}}(\tilde{\chi})$ even flatter.
%\begin{itemize}
  %\item \textcolor{red}{Duality is certainly unexpected. The two theories seem to be very different beasts with no obvious path connecting them.}
 %\item \textcolor{red}{ $\tilde{\Lambda}$, cut off scale of electric theory(UV) and $\tilde{\Lambda}$, cut off scale of magnetic theory(IR) are connected by $\tilde{\Lambda}^{3N-F}\t				ilde{\Lambda}^{3(N-F)-F}=(-1)^{F-N}\mu^F$.}(Serbiology page no 56)
  %\item Adding shift symmetry breaking term of $N_1$ in K$\ddot{\textrm{a}}$hler potential to see if there is any change.
%\end{itemize}

%-----------------------------------------------------------------End of inflation-----------------------------------------------------------------------------------------
%--------------------------------------------------------------------------------------------------------------------------------------------------------------------------
\section{Results}\label{sec:6}
End of inflation occurs when slow roll parameters become unity {\it{i.e.}} $\epsilon,\eta\simeq1$. Solving the equalities
we find inflaton field value at the end of inflation $\chi _{end}\simeq\sqrt{2}M_P$. Now it is visible from 
$V_{\rm{Inf}}(\chi,\sigma)$ and Eq.(\ref{eq:Potder}) that we are left with two free parameters $m$ and 
$\Lambda$ once $p$ is fixed. The value of $\beta$ is taken to be $\mathcal{O}(10^{-3})$  so that 
it satisfies $ \Lambda < m_Q^{\prime} < M_P$.  We have performed a scan over these parameters 
and few of our findings are tabulated in Table \ref{table:tab1}. We find $m$ is mostly restricted by the value of 
curvature perturbation, while $\Lambda$ helps decreasing $r$. We consider $m$ to be below $\Lambda$.
\begin{table}[h]
%\begin{center}
\centering
% $\Lambda$, $m$, and $\chi^{*}$ are provided in terms of $M_P$ unit.}
\begin{tabular}{ | l | l | l | l | l | l |}
    \hline
    $\Lambda $ & $ m $& $\chi^*$ & $r$ & $n_s$  \\ \hline \hline
   %$6\times10^{-6}$ & $5.830 \times 10^{-6}$ & 15.127 & 0.1088 & 0.9657  \\ \hline	
   $8.90\times10^{-4}$ & $5.75\times 10^{-6}$ & 14.95 & 0.099 & 0.965  \\ \hline
%   $7\times10^{-6}$ & $5.720 \times 10^{-6}$ & 14.897 & 0.0966 & 0.9643  \\ \hline
%   $7.8\times10^{-6}$ & $5.530 \times 10^{-6}$ & 14.630 & 0.0832 & 0.9619  \\ \hline
   $1.05\times10^{-3}$ & $5.47 \times 10^{-6}$ & 14.55 & 0.079 & 0.961  \\ \hline
%   $8.2\times10^{-6}$ & $5.400 \times 10^{-6}$ & 14.823 & 0.0748 & 0.9601  \\ \hline
%   $8.5\times10^{-6}$ & $5.255 \times 10^{-6}$ & 14.291 & 0.06739 & 0.9583  \\ \hline
    $1.18\times10^{-3}$ & $4.91 \times 10^{-6}$ & 13.92 & 0.052 & 0.954 \\ \hline
\end{tabular}
\caption{Predictions for $r$, $n_s$ and $\chi^{*}$ are provided for sets of values of parameters 
$m$, $\Lambda$ involved in $V_{\rm{Inf}}$. The dataset corresponds to $N_e=60$, $\alpha=0$, $p=4/7$, $\beta=1.5\times 10^{-3}$ and 
values of ($m$, $\Lambda$, $\chi^*$) are in $M_P$ unit.}
%\label{tab:1}
\label{table:tab1}
\end{table}
Also, we consider effects of non-zero $\alpha$ which is provided in Table \ref{table:tab2}.
\begin{table}[h]
%\caption{For $\alpha=0$, various cosmological parameters, $N_e=60$}
%\end{center}
%\end{table*}
%\begin{table*}
%\centering
\begin{center}
\begin{tabular}{ | l | l | l | l | l | l |}
    \hline
    $\alpha$ & $m$& $\chi^*$ & $r$ & $n_s$  \\ \hline \hline
  %$0.0001$ & $5.440 \times 10^{-6}$ & 14.456 & 0.076 & 0.961  \\ \hline
   $0.0003$ & $5.390 \times 10^{-6}$ & 14.271 & 0.069 & 0.960  \\ \hline	
   $0.0005$ & $5.300 \times 10^{-6}$ & 14.067 & 0.063 & 0.959  \\ \hline
   $0.0007$ & $5.156 \times 10^{-6}$ & 13.841 & 0.055 & 0.957  \\ \hline
  % $0.0009$ & $5.000 \times 10^{-6}$ & 13.607 & 0.048 & 0.955  \\ \hline
  % $0.001$ & $4.900 \times 10^{-6}$ & 13.48  & 0.044 & 0.954  \\ \hline
\end{tabular}
\caption{Predictions for $r$, $n_s$ and $\chi^{*}$ are provided for sets of values of parameters $m$ and $\Lambda$ involved in $V_{\rm{Inf}}$. The dataset corresponds to $N_e=60$, $\Lambda=1.05\times 10^{-3}M_P$, 
$p=4/7$, $\beta=1.5\times 10^{-3}$ and values of ($m$, $\chi^*$) are in $M_P$ unit.}
%\label{tab:2}
\label{table:tab2}
\end{center}
\end{table}

\begin{figure}[h]
\centering
\includegraphics[width=8cm,height=6cm]{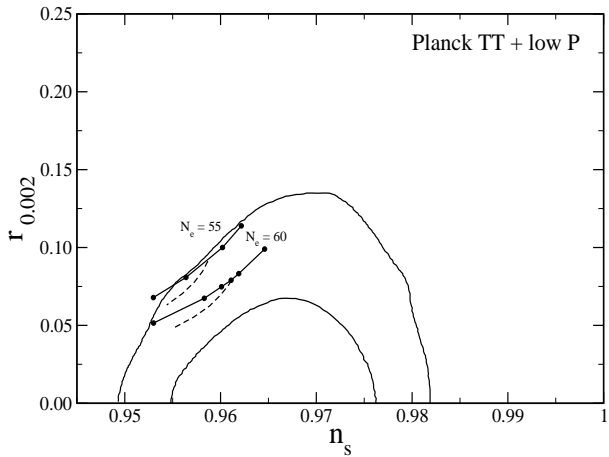}
\caption{Predictions for $n_s$ and $r$ as obtained including dataset from the modified chaotic inflation model (from Table \ref{table:tab1}) indicated by dark dots for $N_e=60$. A solid line joining them represents the prediction 
of $n_s$ and $r$ while $\Lambda$ is varied. Similar predictions for $N_e=55$ are also included. The dashed lines correspond to the predictions by varying $\alpha$  while value of $\Lambda$ is fixed at $1.1\times 10^{-3}M_P$ (for $N_e=55$) and $1.05\times 10^{-3}M_P$ (for $N_e=60$).}
\label{Planck}
\end{figure}

 We find from Table \ref{table:tab1} that corresponding to $\Lambda=1.05\times10^{-3}M_P$, values of $r\sim 0.079$ and $n_s\sim 0.96$ can be achieved with $m\sim5.5\times10^{-6} M_P$. To compare, with the same $\Lambda$, a somewhat lower value of $r\sim0.069$ and $n_s\sim0.96$
are obtained with $m=5.5\times10^{-6}M_P$ and $\alpha=0.
0003$. In obtaining Table \ref{table:tab2}, we have kept $\frac{\alpha\chi^2}{M_P^2}\ll 1$. In Fig.\ref{Planck}, we indicate the respective points of Table \ref{table:tab1} by black dots and note that those points fall within the $2\sigma$ allowed range of  $n_s-r$ plot from PLANCK 2015\cite{Ade:2015xua,Ade:2015lrj} safely. The solid 
line for $N_e=60$ indicates the possible set of points 
(including the ones from Table \ref{table:tab1}) that describe $n_s$ and $r$ for different values of $\Lambda$. Similarly the other solid line corresponds to the set of points for $N_e=55$. The dashed lines describe the effect of introducing $\alpha$. Now we can have an estimate of the mass of the $\delta$ field ($m_{\delta}$) during inflation. 
For $\Lambda=1.05\times10^{-3} M_P$, $m=5.5\times10^{-6} M_P$ we found numerically $\frac{m_{\delta}}{H_{\rm{Inf}}}=\sim 1.2$ during inflation. 
This ensures $\delta$ field to be stabilized at origin. On the other hand, $\frac{m_{\sigma}}{H_{\rm{Inf}}}$ is found to be $\sim 2.5$ which indicates that the fluctuation of $\sigma$-field about $\langle \sigma\rangle$ (in terms of $\chi$) is almost negligible.

\hspace{3mm}

%-----------------------------------------------------------------Dynamics after Inflation----------------------------------------------------------------------------------------
%--------------------------------------------------------------------------------------------------------------------------------------------------------------------------
\section{Dynamics after inflation}\label{sec:dyn2}
Once the inflation is over, the field $\chi$ rolls down along the path as shown in Fig.\ref{fig:Pot1} and $\langle\sigma\rangle$ also follows its VEV which is $\chi$ dependent. 
Note that at the end of inflation, $N_1$ still satisfies $N_1\gg \sqrt{\frac{\Lambda M_P}{\beta}}$ condition. However once $N_1<\Lambda$ is realized, we need to relook into the term responsible for dynamic modification of chaotic inflation. As in this situation $m_Q^{\prime} \ll \Lambda$, the electric quarks $(Q,\tilde{Q})$ can not be integrated out anymore and 
we can use the magnetic dual description of the ISS sector similar to Eq.(\ref{eq:ISS1}) and (\ref{eq:dyn}). Therefore 
the superpotential for the ISS, describing the magnetic dual theory and the RH neutrino becomes
\begin{equation}\label{eq:mag2}
W_m=h{\rm{Tr}}(q\Phi\tilde{q})-h\mu^2{\rm{Tr}}\Phi-\frac{\beta N_1^2{\rm{Tr}}(\Phi)\Lambda}{M_P}+m N_1 N_2.
\end{equation}
To discuss what happens to the $N_1$ and the fields involved in SUSY breaking sector, let us calculate the F-terms as follows
\begin{eqnarray}
F_{\Phi_{ij}}=h\tilde{q}^i_{c}q^{c}_{j}-\Big(h\mu^2+\frac{\beta\Lambda N_1^2}{M_P}\Big)\delta_{ij},\\
F_{q_i}=h\Phi^i_j\tilde{q}^j; \textrm{  }F_{\tilde{q}^j}=h q_i \Phi^i_j,\\
F_{N_1}=-\frac{2\beta\Lambda}{M_P}N_{1}Tr(\Phi)+m N_2,\\
F_{N_2}=m N_1.
\end{eqnarray}
\\

Similar to the original ISS model, here also all the $F$-terms can not be set to zero simultaneously and hence the supersymmetry breaking is realized. The scalar potential becomes
\begin{eqnarray}
V=(N_f-N_c)|hq\tilde{q}-h\mu^{2}-\frac{\beta N_1^2\Lambda}{M_P}|^{2}+N_c|h\mu^{2}+\frac{\beta N_1^2\Lambda}{M_P}|^2\nonumber\\
+|h q\Phi|^2+|h\Phi\tilde{q}|^2+|mN_2-\frac{2\beta N_{1}\rm{Tr}(\Phi)\Lambda}{M_P}|^{2}+m^2|N_1|^2.
\end{eqnarray}
Supergravity corrections are not included in this potential as below the scale $\Lambda$, the SUGRA corrections become negligible. As long as $N_1$ remains nonzero, the minimum of $q$, $\tilde{q}$, $\Phi$ and $N_2$ are given by
\begin{equation}\label{eq:metamin}
\langle q\rangle=\langle \tilde{q} \rangle=\sqrt{\mu^2+\frac{\beta\Lambda N_1^2}{h M_P}}
\left(\begin{array}{l}
\mathbb{1}_{N_f-N_c}\\
0_{N_c}
\end{array}\right),
\textrm{ } \langle\Phi\rangle=0,  \textrm{ }\langle N_2\rangle=0.
\end{equation}
%Now $\frac{\partial V}{\partial \chi}=0$ implies $\langle\chi\rangle=0$,
A point related to $\langle\Phi\rangle$ is pertinent here. In the ISS set-up, a classical flat direction is present in a smaller subspace of $\Phi$ which is essentially lifted by the Coleman–Weinberg (CW)\cite{Intriligator:2006dd} correction and $\langle\Phi\rangle=0$ is achieved. In our set-up there exists a supergravity influenced mass $\sim \frac{m^2N_1^2}{M_P^2}$ for all the components of $\Phi$ once a canonical K$\ddot{\textrm{a}}$hler potential is assumed. This helps $\Phi$ to settle at origin. However once $N_1$ moves to its own minimum which is at $N_1=0$, this induced mass term vanishes and at that stage, CW correction 
becomes important to lift the flatness. For our purpose, we consider $\langle\Phi\rangle$ to be at zero which serves as the local minimum of the theory.
\\

We will now concentrate on the potential involving $N_1$. Assuming all other fields are stabilized at their VEV (with $\langle N_2\rangle=0 \textrm{  as  } \langle\Phi\rangle=0 $) the scalar potential involving $N_1$ becomes
\begin{equation}
V_{N_1}=N_c|h\mu^2+\frac{\beta N_{1}^2\Lambda}{M_P}|^2+m^2|N_1|^2.
\end{equation}
Splitting $N_1$ into real and imaginary components we get
\begin{align}\label{eq:s1}
V_{N_1}(\chi,\eta)=& N_c h^2|\mu|^4+(\chi^4+\eta^4+2\chi^2\eta^2)\frac{N_c\Lambda^2\beta^2}{4M_{P}^2}+\nonumber\\
& \eta^2(\frac{m^2}{2}-\frac{h N_c \beta\mu^2\Lambda}{M_{P}}) +\chi^2(\frac{m^2}{2}+\frac{h N_c\beta\mu^2\Lambda}{M_P}).
\end{align}
By equating $\frac{\partial V(\chi,\eta)}{\partial\eta}$ with zero, we find $\langle\eta\rangle=0$ provided $m^2>\frac{2 h N_c \beta\mu^2\Lambda}{M_P}$. This condition is easily satisfied in our analysis for the allowed range of $m$, $\Lambda$ and $N_c$ with the observation that $\mu$ can be at most $\sim 10^{12}$ GeV for gravity mediated supersymmetry breaking and $h\sim \mathcal{O}(1)$. In case of gauge mediatioin $\mu$ can be even smaller. 
Therefore setting $\eta=0$ Eq.(\ref{eq:s1}) becomes
\begin{align}
V_{\chi}=N_c h^2|\mu|^4+\chi^4\frac{N_c\beta^2\Lambda^2}{4M_{P}^2}+\chi^2(\frac{m^2}{2}+\frac{h N_c\beta\mu^2\Lambda}{M_P}).
\end{align}
It clearly shows that $\chi=0$ is the minimum of the potential with the vacuum energy $V_{0}=N_ch^2|\mu|^4$. So when $N_1$ settles to zero and reheats, the SQCD sector is essentially decoupled as $W_{\rm{Int}}$ vanishes with $N_1=0$. At this stage the ISS sector stands for the supersymmetry breaking in the metastable minima described by Eq.(\ref{eq:meta}) and $\langle\chi\rangle=0$. Reheat will depend on the coupling of $N_1$ with other SM fields.

%-----------------------------------------------------------Dynamical breaking of $U(1)_R$ and gaugino mass---------------------------------------------------------------
%--------------------------------------------------------------------------------------------------------------------------------------------------------------------------
\section{Dynamical breaking of $U(1)_{R}$}\label{sec:R break}
In the construction of the ISS picture of realizing supersymmetry breaking dynamically, $U(1)_R$ symmetry plays an important role. The superpotential $W$ carries  $R$-charge of two units. The $\Phi$ field being linear in the superpotential must also carry the $R$-charge 2 and it is not broken as $\langle\Phi\rangle=0$. A lot of exercises have been performed to achieve $R$-symmetry breaking in order to give mass to the gauginos. One such interesting approach is through the baryon deformation of $W_m$ suggested by \cite{Abel:2007jx}. In \cite{Abel:2007jx} the authors considered the superpotential (for the magnetic theory)
\begin{align}\label{eq:def}
W=\Phi_{ij}q_i\tilde{q}_j-\mu^2\Phi_{ij}+m_{q}\epsilon_{a}^{r}\epsilon_{b}^{s}q^a_{r}q^b_s,
\end{align}
with $N_f=7$ and $N_c=5$, where $r$, $s=1,2$ and $i$, $j=1,..,7$ and $\mu^2=m_Q \Lambda$. R-charges of $q$, $\tilde{q}$ and $\Phi$ are provided in Table \ref{Tab:SQCD} and reason behind this choice is elaborated in Appendix \ref{app:B}. With the specific choice of $N_f$ and $N_c$, the last term is a singlet under the gauge group in the magnetic theory. It represents the baryon deformation, introduction of which shifts the $\langle \Phi\rangle$ to a nonzero value $\sim m_{q}$ and thereby breaking $R$-symmetry spontaneously. In realizing this set-up it was assumed the associated global symmetry $SU(N_f=7)$ is broken down to $SU(5)\times SU(2)$ and the $SU(5)$  after gauging can therefore be  identified with the parent of the Standard Model gauge group. We follow this suggestion for breaking the $U(1)_R$ and argue that this 
approach and the conclusion of \cite{Abel:2007jx} are effectively unaltered by the additional interaction between the SQCD-sector and the inflation sector. In view 
of Eq.(\ref{eq:def}), the charges of $q$, $\tilde{q}$ and $\Phi$ under the discrete symmetries introduced in our framework are provided in Table \ref{Tab:SQCD}.

\begin{table}[h]
\begin{center}
    \begin{tabular}{| l | l | l | l | l |}
\hline
 Fields & $q$ & $\tilde{q}$ & $\Phi$ & $m_Q$   \\ \hline
 $U(1)_R$ & 1 & -1 & 2 & 0  \\ \hline
$\textrm{Z}_2$ &1 &1 &1 & 1      \\ \hline
$\textrm{Z}_4$ & 1& $i$ &$-i$ & $i$      \\ \hline
$\textrm{Z}_4^{\prime}$ &1 &1 & 1 & 1    \\ \hline
\end{tabular}
\caption{$U(1)_R$, $\textrm{Z}_2$, $\textrm{Z}_4$ and $\textrm{Z}_4^{\prime}$ charges of various fields involved in the modified ISS model.}
\label{Tab:SQCD}
\end{center}
\end{table}

%\begin{table*}
%\begin{tabular}{ | l | l | }
%A&B\\
%C&d\\
%\end{tabular}
%\end{table*}

 With the introduction of the additional interaction term ($W_{\rm{Int}}$), we can define an effective $\mu_{eff}$ in the superpotential with $\mu_{eff}^2=\mu^2+\Lambda\frac{N_1^2}{M_P}$. We find the minimal choice as in \cite{Abel:2007jx} $N_f=7$ and $N_c=5$, does not provide enough modification (or flatness) in terms of the inflaton potential. So we have chosen $N_f=9$ and $N_c=7$ so that the gauge group in the magnetic theory remains $SU(2)$ as in \cite{Abel:2007jx}. The global symmetry $SU(9)$ is expected to be broken into $SU(2)\times SU(7)$ explicitly. Taking both these modifications into account, we expect the conclusions of \cite{Abel:2007jx} are essentially remain unchanged, {\it{i.e.}} $\langle\Phi\rangle$ is shifted by an amount $\sim m_q\sim \mathcal{O}(\mu)$ and hence gauginos become massive. The detailed discussion of the $U(1)_R$ breaking is beyond the scope of this paper. Note that this sort of mechanism for breaking $U(1)_R$ holds for $\mu\geq 10^{5}\textrm{ TeV}$
as found in \cite{Abel:2007jx}. The upper limit on $\mu$ can be $\sim10^{12}$ GeV, where gravity mediation dominates over gauge mediation. This range of $\mu$ is consistent in satisfying $m^2>\frac{2N_c\beta\mu^2\Lambda}{M_{P}}$ relation also which keeps the $\langle\eta\rangle$ at origin as discussed in section \ref{sec:dyn2}.

%-------------------------------------------------------------------Neutrino mass and mixing-----------------------------------------------------------------------------
%---------------------------------------------------------------------------------------------------------------------------------------------------------------------------
\section{Neutrino masses and mixing}\label{sec:9}
%At the end of inflation, inflaton started to roll down  towards its minimum. Hubble parameter $H$ decreases with time as $t^{-1}$.
We will discuss reheating and generation of light neutrino masses through the superpotential
\begin{equation}\label{eq:decay1}
W=W_{m}+m_3N_3^2+h_{i\alpha}N_iL_{\alpha}H_u.
\end{equation}
$W_m$ is as described in Eq.(\ref{eq:mag2}). The second and third terms represent the mass term for the third RH neutrino and the neutrino Yukawa couplings with all three RH neutrinos respectively. Note that the superpotential respects the $U(1)_R$ symmetry and therefore the choice of $R$-charges of the $SU(2)_L$ lepton doublets further restricts the Yukawa interaction terms.

\begin{table}[h]
\begin{center}
    \begin{tabular}{ || l | l | l | l | l | l | l | l | l | l | l ||  }
\hline
Fields & $N_1$ & $N_2$ & $N_3$ & $L_1$ & $L_2$ & $L_3$ & $H_{u,d}$ & $z_1$ & $z_2$ & $z_3$  \\ \hline
$U(1)_R$ & $0$  & $2$& $1$ & $2$ & $0$ & $0$ & $\textrm{ }0$ & 0 & 0 & 0  \\ \hline
$\textrm{Z}_2$ & -1 & -1 & 1 & 1 & -1 & -1 & 1 & 1 & 1 & -1\\ \hline
$\textrm{Z}_4$ & 1 & 1 & 1 & 1 & 1 & 1 & 1 & 1 & $i$ & 1\\ \hline
$\textrm{Z}_4^{\prime}$ & 1 & 1 & 1 & $-i$ & 1 & 1 & 1 & 1 & 1 & $i$\\ \hline
\end{tabular}
\caption{$U(1)_R$, $\textrm{Z}_2$, $\textrm{Z}_4$ and $\textrm{Z}^{\prime}_4$ charges of the RH neutrinos, Higgs and Lepton doublets.}
\label{tab:lepton}
\end{center}
\end{table}

With one such typical choice of $R$-charges (only) specified in Table-\ref{tab:lepton}, the allowed Yukawa terms are given by,
%In our case this interaction term will be added with the $W_{mag}$.
%\begin{equation}
%W_{reheat}\supset-\frac{N_1^2 Tr(\Phi) \Lambda}{M_P}+ m N_1 N_2+m^{\prime}N_3^2-\mu^2Tr(\Phi)+h_{i\alpha}N_iL_{\alpha}H_u.
%\end{equation}
%$h_{i\alpha}$ denotes the yukawa coupling of RH neutrino with the lepton doublet $L_{i\alpha}$ and the up-type Higgs $H_u$. Right-handed neutrino mass matrix is given by
%\begin{equation}
%M_{R} =
 %\begin{pmatrix}
  %\langle \Phi \rangle & m & 0 \\
   %m & 0 & 0 \\
   %0 & 0 & m^{\prime}
 %\end{pmatrix}
%\end{equation}
\begin{align}\label{eq:lepto1}
W_Y\supset h_{11}N_1 L_1 H_u + h_{22}N_2 L_2 H_u + h_{23} N_2 L_3 H_u.
\end{align}
%The interaction lagrangian for inflaton(${N_1}$) relevant to its decay during reheating is
%\begin{equation}
%\mathcal{L}\supset\textrm{ }(-2 m N_f\Lambda) N_1 N_2\Phi + m h_{22}N_1H_u\tilde{L_2}+h_{13}N_1\tilde{H}L_3.
%\end{equation}
The coefficient $h_{11}$ can be explained through the vev of another spurion $z_3$ which transforms similar to $z_1$ under 
shift symmetry while odd under the $\textrm{Z}_2$ symmetry considered. A term in the superpotential $(y_1 z_3 N_1 L_1 H_u)/M_P$ then generates $h_{11}=y_1\langle z_3\rangle/M_P$. Here we incorporate another discrete symmetry $\textrm{Z}_4^{\prime}$ under which $z_3$ has charge $i$ and $L_1$ carries $-i$. All the other fields transform trivially under $\textrm{Z}_4^{\prime}$ as seen from Table \ref{tab:lepton}. The new $\textrm{Z}_4^{\prime}$ helps disallow the unwanted terms\footnote{Even with the new $\textrm{Z}_4^{\prime}$, $\frac{(z_3N_1)^4z_1N_1N_2}{M_P^8}$ term will be allowed, however this term turns out to be very small.} like $\frac{(z_3N_1)z_1N_1N_2}{M_P^2}$ and $\frac{z_3 N_1 \textrm{Tr}Q\tilde{Q}}{M_P}$. The superpotential in Eq.(\ref{eq:mag2}) and Eq.(\ref{eq:decay1}) and $W_Y$ in Eq.(\ref{eq:lepto1}) determine the structure of the RH neutrino mass matrix and the Dirac neutrino mass matrix as
\begin{align}
M_{R}= \begin{pmatrix}
\varepsilon_m& m & 0 \\
m & 0 & 0 \\
0 & 0 & m_{3}
\end{pmatrix}
\textrm{ ; }
m_{D}=\langle H_u \rangle \begin{pmatrix}\label{eq:Dirac mat}
h_{11} & 0 & 0 \\
0 & h_{22} & h_{23} \\
0 & 0 & 0
\end{pmatrix},
\end{align}

with $\varepsilon_m=\frac{\beta\langle\Phi\rangle\Lambda}{M_P}\ll m$. Here we have incorporated the $\langle \Phi\rangle$ related to the deformation as discussed in section \ref{sec:R break}.

 Light neutrino mass-matrix can therefore be obtained from the type-I seesaw\cite{Mohapatra:1979ia}
contribution $m_\nu=m_D^T\frac{1}{M_R}m_D$ and is given by
\begin{equation}\label{eq:lepto3}
m_{\nu} =\frac{\langle H_u\rangle^2}{m}
 \begin{pmatrix}
   0 & h_{11}h_{22} & h_{11}h_{23} \\
   h_{11}h_{22} & -\frac{\varepsilon_m h_{22}^2}{m} & -\frac{\varepsilon_m h_{22} h_{23}}{m} \\
   h_{11}h_{23} & -\frac{\varepsilon_m h_{22}h_{23}}{m} & -\frac{\varepsilon_m h_{22}^2}{m}
 \end{pmatrix}.
\end{equation}
Note that all the terms involving $\varepsilon_m/m$ are much smaller compared to the 12(21) and 13(31) entries of $m_{\nu}$. Once the terms proportional to $\varepsilon_m/m$ are set to zero, $m_\nu$ coincides with the neutrino mass matrix proposed in \cite{Barbieri:1998mq} leading to an inverted hierarchical spectrum of light neutrinos. The above texture of $m_{\nu}$ in Eq.(\ref{eq:lepto3}) then predicts
\begin{align}\label{eq:neutrino}
m_{\nu_1}\simeq m_{\nu_2}\simeq\sqrt{2}\frac{\kappa h_{11} v_u^2}{m} ;\textrm{  } m_{\nu_3}\simeq\frac{\kappa^2v_u^2}{2m}\Big(\frac{\varepsilon_m}{m} \Big), \textrm{  }\\
 \Delta m_{12}^2\simeq\frac{\kappa^3 h _{11}v_u^4}{m^2}\Big(\frac{\varepsilon_m}{m} \Big) \textrm{ and } \Delta m^2_{23}\simeq\frac{2 \kappa^2 h_{11}^2v_u^4}{m^2},
\end{align}
where $h_{22}\simeq h_{23}= \kappa$ is assumed for simplicity and $\langle H_u \rangle=v_u$. It also indicates a bi-maximal mixing pattern in solar and atmospheric sectors along with $\theta_{13}\simeq\frac{\varepsilon_m}{m}\frac{\kappa}{h_{11}}$.

Now as $m$ is essentially determined from the inflation part in our scenario, we find $h_{11}$ of order $\sim \mathcal{O}(10^{-2})$ to get correct magnitude of $\Delta m_{23}^2\simeq 2.5\times 10^{-3}\textrm{ eV}^2$\cite{Agashe:2014kda}  with $v_u=174$ GeV and $\kappa=1$ . At first sight it is tantalizing to note that with $\frac{\varepsilon_m}{m}\ll1$ and we could also accommodate $\Delta m_{12}^2$ ($\simeq 7\times 10^{-5}\textrm{ eV}^2$\cite{Agashe:2014kda}). However $\frac{\varepsilon_m}{m}\simeq \frac{\beta\Lambda}{m}\frac{\rm{Tr\langle\Phi\rangle}}{M_P}\sim \frac{\mu}{M_P}\leq\mathcal{O}(10^{-6})$ and it turns out to be too small (a value of $\frac{\varepsilon_m}{m}\sim 10^{-2}$ could fulfill the requirement) to explain the solar splitting correctly. Therefore small but relatively larger entries are required in place of $\Big(\frac{\varepsilon_m}{m}\Big)$ terms in $m_{\nu}$\cite{Babu:2002ex}. A possible source of these terms could arise in our case from higher order R-symmetry breaking terms. The mixing angles $\theta_{12},\textrm{}\
\theta_{13}$ can be corrected from the contribution in the charged lepton sector. We do not explore this possibility in detail here. It could as well be the effect of renormalization group evolutions as pointed out by  \cite{Babu:2002ex}, or even other sources ({\it{e.g.}} type-II contribution as in \cite{Karmakar:2015jza}) of neutrino mass.
%For example, the following superpotential in the charged lepton
%\begin{align}
%W_{e}=x_{13}\bar{e}_1 L_3 H_d+x_{22}\bar{e}_2 L_2 H_d + %x_{33}\bar{e}_3 L_3 H_d,
%\end{align}
%produces the charged lepton mass matrix as
%\begin{equation}
%m_{L} =x_{33}{\langle H_d \rangle^2}
% \begin{pmatrix}
%   0 & 0 & \frac{x_{13}}{x_{33}} \\
%   0 & \frac{x_{22}}{x_{33}} & 0 \\
%   0 & 0 & 1
% \end{pmatrix},
%\end{equation}
%Then a small contribution in 31 entry can help in correcting %$\theta_{12}$, realizing $\theta_{13}$ to an acceptable level %\cite{Agashe:2014kda}.
%------------------------------------------------------------------------------------------------------------------------------------------------------------------------------------------------------------------
%-----------------------------------------------------------------------------------------------------------------------------------------------------------------------------------------------------------------

\section{Reheating}\label{sec:10}
 As soon as Hubble parameter becomes less than the mass of the inflaton, $N_1$ starts to oscillate around its minimum and universe reheats. 
The estimate of $h$ helps us determining the reheat temperature. The decay of $N_1$ is governed through the $W$ in Eq.(\ref{eq:decay1}). The decay width therefore is estimated to be
\begin{equation}
\Gamma_{N_1}= \frac{(2\kappa^2+h_{11}^2)}{8\pi} m,
\end{equation}
neglecting the effect of $\varepsilon_m$ term. The corresponding reheat temperature is obtained as
\begin{equation}
T_{RH}=\Big(\frac{45}{2\pi^2g_{*}}\Big)^{1/4}\sqrt{\Gamma_{N_1}M_P}\simeq 4\times10^{14}\textrm{ GeV},
\end{equation}
where $m\sim 10^{-6} M_P$ is considered and $\kappa\sim\mathcal{O}(1)$, as obtained from the discussion of the previous section. Such a high reheating temperature poses a threat in terms of over abundance of thermally produced gravitinos\footnote{Note that the chaotic inflation is free from gravitino problem indeed for the non-thermal decay of inflaton\cite{Kawasaki:2006hm,Kawasaki:2006gs}.}. Their abundance is mostly proportional to the reheat temperature \cite{Kawasaki:2008qe},
\begin{equation}
Y_{3/2}\simeq2\times 10^{-9}\Big(\frac{T_{RH}}{10^{13}\textrm{ GeV}}\Big),
\end{equation}
where $Y_{3/2}=\frac{n_{3/2}}{s}$ with $n_{3/2}$ as the number density of gravitinos and $s$ is the entropy density. These gravitinos, if massive, then decays into the lightest supersymmetric particles (LSP) and can destroy the predictions of primordial abundance of light elements. On the other hand, if gravitino is the LSP, the reheating temperature can not be as high as mentioned in our work.
This problem can be circumvented if the gravitinos are superlight, {\it{e.g.}} $m_{3/2}\sim 16 \textrm{ eV}$\cite{Viel:2005qj}. Such a gravitino can be accommodated in the gauge mediated supersymmetry breaking. In our set-up, $\mu$ is the scale which in turn predicts the gravitino mass through $m_{3/2}\simeq \frac{\mu^2}{\sqrt{3}M_P}$. Therefore with $\mu\sim 10^5 \textrm{ GeV}$, such a light gravitino mass can be obtained. Another way to circumvent this gravitino problem is through the late time entropy production\cite{Kohri:2004qu}. Apart from these possibilities one interesting observation by \cite{Fukushima:2013vxa} could be of help in this regard. The author in \cite{Fukushima:2013vxa} have shown that once the messenger mass scale (in case of gauge mediation of supersymmetry breaking) falls below the reheat temperature, the relic abundance of thermally produced gravitinos becomes insensitive to $T_{RH}$ and a large $T_{RH}\sim 10^{13-14}$ GeV can be realized.

Finally we make  brief comments on leptogenesis in the present scenario. Considering $m_3\ll m$, $N_3$ would contribute mostly for the lepton asymmetry production. The CP asymmetry generated can be estimated as \cite{Buchmuller:1998zf}
\begin{equation}
\epsilon _3=\frac{3}{8\pi v_u^2}\frac{1}{(\hat{m}_D^{\dagger}\hat{m}_D)_{33}}\sum_{i=1,2}\textrm{Im}[(\hat{m}_D^{\dagger}\hat{m}_D)^2_{i3}]\frac{m_3}{m}.
\end{equation}
Here $\hat{m_D}$ represents the rotated Dirac mass matrix in the basis where $M_R$ is diagonal. It is found that CP-asymmetry  exactly vanishes in this case. We expect this can be cured with the introduction of higher order $U(1)_R$ symmetry breaking terms which could be introduced into $m_D$ and $M_R$\footnote{We have already mentioned about this possibility of inclusion of such (small) term in the previous section, that can correct the $\Delta m_{12}^2$ and the lepton mixing angles.}. Then similar to \cite{Jenkins:2008rb}, a non-zero lepton asymmetry through the decay of $N_3$ can be realized.
%So total decay width of $N_1$ is\footnote{Note that $N_2$ being as heavy as $N_1$, $N_1$ decay into $\Phi$ and $N_2$ is kinematically forbidden.}
%---------------------------------------------------------------Conclusion----------------------------------------------------------------------------------------------
%------------------------------------------------------------------------------------------------------------------------------------------------------------------------
\section{Conclusion}\label{sec:conclude}
We have considered the superpartner of a right-handed neutrino as playing the role of inflaton. Although a minimal chaotic inflation scenario out of this consideration is a well studied subject, its simplest form is almost outside the 2$\sigma$ region of recent $n_s-r$ plot by PLANCK 2015. We have shown in this work, that a mere coupling with the SQCD sector responsible for supersymmetry breaking can be considered as a deformation to the minimal version of the chaotic inflation. Such a deformation results in a successful entry of the chaotic inflation into the latest $n_s-r$ plot. Apart from this, the construction also ensures that a remnant supersymmetry breaking is realized at the end of inflation. The global $U(1)_R$ symmetry plays important role in constructing the superpotential for both the RH neutrino as well as SQCD sector. We have shown that the shift-symmetry breaking terms in the set up can be accommodated in an elegant way by introducing spurions. Their introduction, although ad hoc, can not only explain the size of the symmetry breaking but also provide a prescription for operators involving the RH neutrino superfields (responsible for inflation) in the superpotential. With the help of the $R$-symmetry and the discrete symmetries introduced, we are able to show that light neutrino masses and mixing resulted from the set-up can accommodate the recent available data nicely,
predicting an inverted hierarchy for light neutrinos. However there still exists a scope for further study in terms of leptogenesis through the $R$-symmetry breaking terms.
%--------------------------------------------------------------------------------------------------------------------------------------------------------------------------
%-----------------------------------------------------------------------Appendix-------------------------------------------------------------------------------------------
\\
\vskip 1cm 
\textbf{\LARGE{Appendix}}
\appendix
%\begin{appendices}
\section{Finding the root of $\sigma$.}\label{App:Appendix1}
Setting $\frac{\partial V_{\rm{Inf}}(\tilde{\chi},\tilde{\sigma})}{\partial\tilde{\sigma}}=0$, we get a fifth order polynomial equation in $\tilde{\sigma}$ of the form,
\begin{align}\label{eq:poly}
\frac{\tilde{\sigma}^5}{2}+\tilde{\sigma}^{3}+k_1\Big[1-\frac{6(1-p)}{\tilde{\chi}^2}+\tilde{\sigma}^2\Big\{ \frac{1}{4}-\frac{6(1-p)}{\tilde{\chi}^2}\Big\}-\frac{\tilde{\sigma}^4}{4}\Big]+k_2\tilde{\sigma}=0,
\end{align}
where $k_1=-\frac{2A}{\tilde{m}^2}\tilde{\chi}^{5-6p}$ and $k_2=\frac{4}{\tilde{\chi}^2}\Big[1+\tilde{\chi}^{10-12p}
\frac{A^2}{2\tilde{m}^4}\Big\{9(1-p)^2-\frac{\tilde{\chi}^2}{4}\Big\}\Big]$. 
Here we disregard the first and third terms from the coefficient of $\tilde{\sigma}^3$ in 
Eq.(\ref{eq:Potder}) as $\tilde{\chi}$ being greater than one during inflation, 
$\frac{\tilde{m}^2\tilde{\chi}^2}{4}$ is the dominant contribution. We 
now try to solve the Eq.(\ref{eq:poly}) to express $\langle\tilde{\sigma}\rangle$ 
in terms of $\tilde{\chi}$. In doing so, note that $\tilde{\chi}$ being inflaton 
is super-Planckian while $\tilde{\sigma}$ remains sub-Planckian ($\tilde{\sigma}<1$) 
during inflation. Also the parameters involved, $\tilde{\Lambda}$ and $\tilde{m}$, 
are considered to be much less than one (in $M_P$ unit), $\tilde{\Lambda}$, 
$\tilde{m}\ll 1$ with $\tilde{\Lambda}\tilde{\chi}\ll 1$. We have also taken 
$\tilde{\Lambda}\geq \tilde{m}$. Since the added contribution via 
$W_{\rm{Int}}$ is expected to provide modification only on the minimal chaotic inflation, 
it is natural that $m$ should be close to $10^{13}$ GeV (also $\tilde{\chi}$ is expected of 
order $\mathcal{O}(10)$). These consideration keeps $k_1$ to be less than one ($k_1<1$) although $k_2$ can be somewhat larger.

\begin{figure}[h]
\hspace{.5cm}
\centering
\includegraphics[width=7cm]{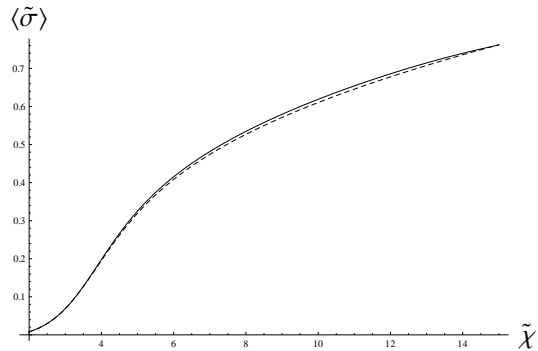}
\caption{Comparison of $\langle\tilde{\sigma}\rangle$ with $\tilde{\chi}$ using perturbation (solid line) and exact numerical result (dashed line). }
\label{fig:3}
\end{figure}

With $p=4/7$, we find $\tilde{\sigma}^4$ can be neglected and the Eq.(\ref{eq:poly}) then reduces to the form
\begin{align}\label{eq:poly1}
\frac{\tilde{\sigma}^5}{2}+\Big( \tilde{\sigma}^3+c_1\tilde{\sigma}^2+k_2\tilde{\sigma}+c_3\Big)=0,
\end{align}
 where $c_1=k_1\Big(\frac{1}{4}-\frac{18}{7\tilde{\chi}^2}\Big)$ and $c_3=k_1\Big(1-\frac{18}{7\tilde{\chi}^2}\Big)$. The coefficient of $\tilde{\sigma}^5$ being $1/2$, the $\tilde{\sigma}^5$ term can be considered as a perturbation over the cubic equation in $\tilde{\sigma}$, as indicated by the first brackets in Eq.(\ref{eq:poly1}). Let $\tilde{\sigma_0}$ be the solution of this cubic part of Eq.(\ref{eq:poly1}) and the analytic form of it can easily be obtained (for real root). Then we consider the solution of Eq.(\ref{eq:poly1}) as
 \begin{align}
 \tilde{\sigma}=\tilde{\sigma}_0+\epsilon\tilde{\sigma}_1+
 \epsilon^2\tilde{\sigma}_2+\epsilon^3\tilde{\sigma}_3,
 \end{align}
  with $\epsilon=1/2$ (coefficient of $\tilde{\sigma}_5$ term) as a perturbation parameter. Finally we get
\begin{align}
\tilde{\sigma}_1 &=\frac{-\tilde{\sigma}_0^5}{k_2+2c_1\tilde{\sigma}_0+3\tilde{\sigma}_0^2},\\
\tilde{\sigma}_2 &=\frac{-5\tilde{\sigma}_0^4\tilde{\sigma}_1-3\tilde{\sigma}_0\tilde{\sigma}_1^2-c_1\tilde{\sigma}_1^2}{k_2+2c_1\tilde{\sigma}_0+3\tilde{\sigma}_0^2},\\
\tilde{\sigma}_3 &=\frac{-5\tilde{\sigma}_0^4\tilde{\sigma}_2-10\tilde{\sigma}_0^3\tilde{\sigma}_1^2-6\tilde{\sigma}_0\tilde{\sigma}_1\tilde{\sigma}_2-\tilde{\sigma}_1^3-2c_1\tilde{\sigma}_1\tilde{\sigma}_2}{k_2+2c_1\tilde{\sigma}_0+3\tilde{\sigma}_0^2}.
\end{align}
We have checked numerically (using mathematica) that this perturbation method for solving the fifth order polynomial equation as in Eq.(\ref{eq:poly}) works reasonably well.  For comparison, we have included Fig.\ref{fig:3} where $\langle\tilde{\sigma}\rangle$ is depicted against the variation of $\tilde{\chi}$ (particularly during inflation when $\chi$ acquires super-Planckian value). The solid line represents the VEV of $\tilde{\sigma}$ as obtained from our perturbation method and the dashed line gives the exact numerical estimate of $\langle\tilde{\sigma}\rangle$ from Eq.(\ref{eq:poly}). In order to get $V_{\rm{Inf}}$ in terms of $\tilde{\chi}$, we have used the analytic form of $\langle\tilde{\sigma}\rangle$ obtained through this perturbation method.
%---------------------------------------------------------------------Charge Distribution of various fields----------------------------------------------------------------
%--------------------------------------------------------------------------------------------------------------------------------------------------------------------------
\section{$R$ charges of various fields}\label{app:B}
Here we discuss the $R$-charge assignments for the various fields involved in our construction. Firstly in Table \ref{tab:app1}, we include various $U(1)$ global charges associated with massless SQCD theory ($N_f=9$, $N_c=7$) following\cite{Intriligator:2007cp}.
\begin{table}[h]
\begin{center}
    \begin{tabular}{ | l | l | l |l|}
    \hline
   Fields& $U(1)_B$ &  $U(1)_A$ & $U(1)_{R^{\prime}}$ \\ \hline \hline	
   Q& 1& 1& $\frac{2}{9}$ \\
  $ \tilde{Q}$ & -1& 1&   $\frac{2}{9}$    \\ \hline
  $\Lambda$ & 0 & $\frac{3}{2}$ & 0  \\
  $W$ & 0 & 0 & 2 \\ \hline
  $q$ & $\frac{7}{2}$ & -$\frac{1}{4}$ & $\frac{7}{9}$ \\
  $\tilde{q}$ &  -$\frac{7}{2}$ & -$\frac{1}{4}$ & $\frac{7}{9}$  \\
  $\Phi$ & $ 0 $ & $\frac{1}{2}$ & $\frac{4}{9}$ \\  \hline

\end{tabular}
\caption{Global charges of various fields in a massless SQCD theory\cite{Intriligator:2007cp}.}
\label{tab:app1}
\end{center}
\end{table}
 However once the term $m_Q\textrm{Tr}Q\tilde{Q}$
is included in the UV description and a baryonic deformation (through $m_{q}qq$ term in Eq.(\ref{eq:def})) is considered as well in the magnetic description, there exists a residual $U(1)_R$ symmetry only. The charges of the fields in the magnetic description can be obtained \cite{Durnford:2009aba} from
\begin{align}
R=\frac{2}{7}B+\frac{28}{9}A+R^{\prime}.		
\end{align}
This redefined R-charges are mentioned in Table \ref{Tab:SQCD}. The superpotential in Eq.(\ref{eq:decay1}) respects this $U(1)_R$ symmetry. 
From $\Phi$  = Tr$(Q\tilde{Q})/ \Lambda$,  the $Q\tilde{Q}$ combination has two units of $R$ charges.

%\begin{table}
%\begin{center}
 %   \begin{tabular}{ || l | l | l | l | l | l | l | l | l | l | l || }
%\hline
%Fields & $N_1$ & $N_2$ & $N_3$ & $L_1$ & $L_2$ & $L_3$ & $H_{u,d}$ & $\bar{e}_1$ & $\bar{e}_2$ & $\bar{e}_3$ \\ \hline
%$U(1)_R$ &-$\frac{7}{3}$  & $\frac{13}{3}$& $1$ & $\frac{13}{3}$ & $-\frac{7}{3}$& -$\frac{-7}{3}$ & $0$ & $\frac{13}{3}$ & $\frac{13}{3}$ & %$\frac{13}{3}$ \\ \hline
%\end{tabular}
%\caption{$U(1)_R$ charges of lepton sector.}
%\label{tab:app3}
%\end{center}
%\end{table}



\begin{thebibliography}{99}

%\cite{Ade:2014xna}
\bibitem{Ade:2014xna}
  P.~A.~R.~Ade {\it et al.}  [BICEP2 Collaboration],
  %``Detection of $B$-Mode Polarization at Degree Angular Scales by BICEP2,''
  Phys.\ Rev.\ Lett.\  {\bf 112}, no. 24, 241101 (2014)
  [arXiv:1403.3985 [astro-ph.CO]].
  %%CITATION = ARXIV:1403.3985;%%
  %1066 citations counted in INSPIRE as of 16 juin 2015
\bibitem{Ade:2015xua}
  P.~A.~R.~Ade {\it et al.}  [Planck Collaboration],
  %``Planck 2015 results. XIII. Cosmological parameters,''
  arXiv:1502.01589 [astro-ph.CO].
  %%CITATION = ARXIV:1502.01589;%%
  %329 citations counted in INSPIRE as of 16 juin 2015
%\cite{Ade:2015lrj}
\bibitem{Ade:2015lrj}
  P.~A.~R.~Ade {\it et al.}  [Planck Collaboration],
  %``Planck 2015 results. XX. Constraints on inflation,''
  arXiv:1502.02114 [astro-ph.CO].
  %%CITATION = ARXIV:1502.02114;%%
  %156 citations counted in INSPIRE as of 16 juin 2015
%\cite{Kawasaki:2000yn}
\bibitem{Kawasaki:2000yn}
  M.~Kawasaki, M.~Yamaguchi and T.~Yanagida,
  %``Natural chaotic inflation in supergravity,''
  Phys.\ Rev.\ Lett.\  {\bf 85}, 3572 (2000)
  [hep-ph/0004243].
  %%CITATION = HEP-PH/0004243;%%
  %289 citations counted in INSPIRE as of 17 juin 2015
%\cite{Kawasaki:2000ws}
\bibitem{Kawasaki:2000ws}
  M.~Kawasaki, M.~Yamaguchi and T.~Yanagida,
  %``Natural chaotic inflation in supergravity and leptogenesis,''
  Phys.\ Rev.\ D {\bf 63}, 103514 (2001)
  [hep-ph/0011104].
  %%CITATION = HEP-PH/0011104;%%
  %80 citations counted in INSPIRE as of 23 juin 2015
%\cite{Linde:2007fr}
\bibitem{Linde:2007fr}
  A.~D.~Linde,
  %``Inflationary Cosmology,''
  Lect.\ Notes Phys.\  {\bf 738}, 1 (2008)
  [arXiv:0705.0164 [hep-th]].
  %%CITATION = ARXIV:0705.0164;%%
  %297 citations counted in INSPIRE as of 27 juil. 2015
%\cite{Harigaya:2014fca}
\bibitem{Harigaya:2014fca}
  K.~Harigaya, M.~Kawasaki and T.~T.~Yanagida,
  %``Lower bound of the tensor-to-scalar ratio $r \mathop{}_{\textstyle \sim}^{\textstyle >} 0.1$ in a nearly quadratic chaotic inflation model in supergravity,''
  Phys.\ Lett.\ B {\bf 741}, 267 (2015)
  [arXiv:1410.7163 [hep-ph]].
  %%CITATION = ARXIV:1410.7163;%%
  %1 citations counted in INSPIRE as of 24 Apr 2015
%\cite{Li:2013nfa}
\bibitem{Li:2013nfa}
  T.~Li, Z.~Li and D.~V.~Nanopoulos,
  %``Supergravity Inflation with Broken Shift Symmetry and Large Tensor-to-Scalar Ratio,''
  JCAP {\bf 1402}, 028 (2014)
  [arXiv:1311.6770 [hep-ph]].
  %%CITATION = ARXIV:1311.6770;%%
  %23 citations counted in INSPIRE as of 07 juil. 2015
%\cite{Ade:2013zuv}
\bibitem{Ade:2013zuv}
  P.~A.~R.~Ade {\it et al.} [Planck Collaboration],
  %``Planck 2013 results. XVI. Cosmological parameters,''
  Astron.\ Astrophys.\  {\bf 571}, A16 (2014)
  [arXiv:1303.5076 [astro-ph.CO]].
  %%CITATION = ARXIV:1303.5076;%%
  %3684 citations counted in INSPIRE as of 29 Jul 2015
%\cite{Nakayama:2013txa}
\bibitem{Nakayama:2013txa}
  K.~Nakayama, F.~Takahashi and T.~T.~Yanagida,
  %``Polynomial Chaotic Inflation in Supergravity,''
  JCAP {\bf 1308}, 038 (2013)
  [arXiv:1305.5099 [hep-ph]].
  %%CITATION = ARXIV:1305.5099;%%
  %44 citations counted in INSPIRE as of 27 juil. 2015
%\cite{Nakayama:2014wpa}
\bibitem{Nakayama:2014wpa}
  K.~Nakayama, F.~Takahashi and T.~T.~Yanagida,
  %``Polynomial Chaotic Inflation in Supergravity Revisited,''
  Phys.\ Lett.\ B {\bf 737}, 151 (2014)
  [arXiv:1407.7082 [hep-ph]].
  %%CITATION = ARXIV:1407.7082;%%
  %3 citations counted in INSPIRE as of 27 juil. 2015
%\cite{Murayama:1992ua}
\bibitem{Murayama:1992ua}
  H.~Murayama, H.~Suzuki, T.~Yanagida and J.~Yokoyama,
  %``Chaotic inflation and baryogenesis by right-handed sneutrinos,''
  Phys.\ Rev.\ Lett.\  {\bf 70}, 1912 (1993).
  %%CITATION = PRLTA,70,1912;%%
  %207 citations counted in INSPIRE as of 01 juil. 2015
%\cite{Ellis:2004hy}
\bibitem{Ellis:2004hy}
  J.~R.~Ellis,
  %``Sneutrino inflation,''
  Nucl.\ Phys.\ Proc.\ Suppl.\  {\bf 137}, 190 (2004)
  [hep-ph/0403247].
  %%CITATION = HEP-PH/0403247;%%
  %6 citations counted in INSPIRE as of 01 juil. 2015
    %\cite{Khalil:2011kd}
\bibitem{Khalil:2011kd}
  S.~Khalil and A.~Sil,
  %``Right-handed Sneutrino Inflation in SUSY B-L with Inverse Seesaw,''
  Phys.\ Rev.\ D {\bf 84}, 103511 (2011)
  [arXiv:1108.1973 [hep-ph]].
  %%CITATION = ARXIV:1108.1973;%%
  %3 citations counted in INSPIRE as of 25 août 2015
%\cite{Murayama:2014saa}
\bibitem{Murayama:2014saa}
  H.~Murayama, K.~Nakayama, F.~Takahashi and T.~T.~Yanagida,
  %``Sneutrino Chaotic Inflation and Landscape,''
  Phys.\ Lett.\ B {\bf 738}, 196 (2014)
  [arXiv:1404.3857 [hep-ph]].
  %%CITATION = ARXIV:1404.3857;%%
  %18 citations counted in INSPIRE as of 16 Jun 2015
%\cite{Evans:2015mta}
\bibitem{Evans:2015mta}
  J.~L.~Evans, T.~Gherghetta and M.~Peloso,
  %``Affleck-Dine Sneutrino Inflation,''
  Phys.\ Rev.\ D {\bf 92}, no. 2, 021303 (2015)
  [arXiv:1501.06560 [hep-ph]].
  %%CITATION = ARXIV:1501.06560;%%
  %2 citations counted in INSPIRE as of 27 juil. 2015
%\cite{Intriligator:2006dd}
\bibitem{Intriligator:2006dd}
  K.~A.~Intriligator, N.~Seiberg and D.~Shih,
  %``Dynamical SUSY breaking in meta-stable vacua,''
  JHEP {\bf 0604}, 021 (2006)
  [hep-th/0602239].
  %%CITATION = HEP-TH/0602239;%%
  %531 citations counted in INSPIRE as of 24 Apr 2015
  %\cite{Brax:2008rk}
\bibitem{Brax:2008rk}
  P.~Brax, C.~A.~Savoy and A.~Sil,
  %``Intermediate Scale Inflation and Metastable Supersymmetry Breaking,''
  Phys.\ Lett.\ B {\bf 671}, 374 (2009)
  [arXiv:0807.1569 [hep-ph]].
  %%CITATION = ARXIV:0807.1569;%%
  %1 citations counted in INSPIRE as of 19 juin 2015
%\cite{Savoy:2007jb}
\bibitem{Savoy:2007jb}
  C.~A.~Savoy and A.~Sil,
  %``Can Inflation Induce Supersymmetry Breaking in a Metastable Vacuum?,''
  Phys.\ Lett.\ B {\bf 660}, 236 (2008)
  [arXiv:0709.1923 [hep-ph]].
  %%CITATION = ARXIV:0709.1923;%%
  %4 citations counted in INSPIRE as of 19 Jun 2015
%\cite{Brax:2009yd}
\bibitem{Brax:2009yd}
  P.~Brax, C.~A.~Savoy and A.~Sil,
  %``SQCD Inflation & SUSY Breaking,''
  JHEP {\bf 0904}, 092 (2009)
  [arXiv:0902.0972 [hep-ph]].
  %%CITATION = ARXIV:0902.0972;%%
  %4 citations counted in INSPIRE as of 19 Jun 2015
%\cite{Craig:2008tv}
\bibitem{Craig:2008tv}
  N.~J.~Craig,
  %``ISS-flation,''
  JHEP {\bf 0802}, 059 (2008)
  [arXiv:0801.2157 [hep-th]].
  %%CITATION = ARXIV:0801.2157;%%
  %8 citations counted in INSPIRE as of 19 juin 2015
%\cite{Abel:2007jx}
\bibitem{Abel:2007jx}
  S.~Abel, C.~Durnford, J.~Jaeckel and V.~V.~Khoze,
  %``Dynamical breaking of U(1)(R) and supersymmetry in a metastable vacuum,''
  Phys.\ Lett.\ B {\bf 661}, 201 (2008)
  [arXiv:0707.2958 [hep-ph]].
  %%CITATION = ARXIV:0707.2958;%%
  %91 citations counted in INSPIRE as of 24 Apr 2015

%\cite{Hooft:1980}
 \bibitem{Hooft:1980}
 G. $^{\textrm{,}}$t Hooft, in {\it{Recent Developments in Gauge Theories}}, edited by G. $^{\textrm{,}}$t Hooftet al. (Plenum, Carg$'$ese, 1980).

%\cite{Harigaya:2015pea}
\bibitem{Harigaya:2015pea}
  K.~Harigaya, M.~Ibe, M.~Kawasaki and T.~T.~Yanagida,
  %``Revisiting the Minimal Chaotic Inflation Model,''
  arXiv:1506.05250 [hep-ph].
  %%CITATION = ARXIV:1506.05250;%%
  %1 citations counted in INSPIRE as of 27 juil. 2015
%\cite{Pallis:2015mga}
\bibitem{Pallis:2015mga}
  C.~Pallis,
  %``Kinetically modified nonminimal chaotic inflation,''
  Phys.\ Rev.\ D {\bf 91}, no. 12, 123508 (2015)
  [arXiv:1503.05887 [hep-ph]].
  %%CITATION = ARXIV:1503.05887;%%
%\cite{Barenboim:2015lla}
\bibitem{Barenboim:2015lla}
  G.~Barenboim and W.~I.~Park,
  %``New- vs. chaotic-inflations,''
  arXiv:1504.02080 [astro-ph.CO].
  %%CITATION = ARXIV:1504.02080;%%
  %1 citations counted in INSPIRE as of 27 juil. 2015
%\cite{Carpenter:2014saa}
\bibitem{Carpenter:2014saa}
  L.~M.~Carpenter and S.~Raby,
  %``Chaotic hybrid inflation with a gauged B–L,''
  Phys.\ Lett.\ B {\bf 738}, 109 (2014)
  [arXiv:1405.6143 [hep-ph]].
  %%CITATION = ARXIV:1405.6143;%%
  %3 citations counted in INSPIRE as of 27 juil. 2015
  %\cite{Heurtier:2015ima}
\bibitem{Heurtier:2015ima}
  L.~Heurtier, S.~Khalil and A.~Moursy,
  %``Single Field Inflation in Supergravity with a $U(1)$ Gauge Symmetry,''
  arXiv:1505.07366 [hep-ph].
  %%CITATION = ARXIV:1505.07366;%%
%\cite{Buchmuller:2014pla}
\bibitem{Buchmuller:2014pla} 
  W.~Buchmuller, E.~Dudas, L.~Heurtier and C.~Wieck,
  %``Large-Field Inflation and Supersymmetry Breaking,''
  JHEP {\bf 1409}, 053 (2014)
  [arXiv:1407.0253 [hep-th]].
  %%CITATION = ARXIV:1407.0253;%%
  %16 citations counted in INSPIRE as of 31 août 2015
%\cite{Intriligator:2007cp}
\bibitem{Intriligator:2007cp}
  K.~A.~Intriligator and N.~Seiberg,
  %``Lectures on Supersymmetry Breaking,''
  Class.\ Quant.\ Grav.\  {\bf 24}, S741 (2007)
  [hep-ph/0702069].
  %%CITATION = HEP-PH/0702069;%%
  %170 citations counted in INSPIRE as of 24 Apr 2015
%\cite{Harigaya:2014wta}
\bibitem{Harigaya:2014wta}
  K.~Harigaya, M.~Ibe, K.~Schmitz and T.~T.~Yanagida,
  %``Dynamical fractional chaotic inflation,''
  Phys.\ Rev.\ D {\bf 90}, no. 12, 123524 (2014)
  [arXiv:1407.3084 [hep-ph]].
  %%CITATION = ARXIV:1407.3084;%%
  %3 citations counted in INSPIRE as of 29 juil. 2015
%\cite{Harigaya:2014sua}
\bibitem{Harigaya:2014sua}
  K.~Harigaya, M.~Ibe, K.~Schmitz and T.~T.~Yanagida,
  %``Dynamical Chaotic Inflation in the Light of BICEP2,''
  Phys.\ Lett.\ B {\bf 733}, 283 (2014)
  [arXiv:1403.4536 [hep-ph]].
  %%CITATION = ARXIV:1403.4536;%%
  %32 citations counted in INSPIRE as of 29 juil. 2015
%\cite{Gao:2014fha}
\bibitem{Gao:2014fha}
  X.~Gao, T.~Li and P.~Shukla,
  %``Fractional chaotic inflation in the lights of PLANCK and BICEP2,''
  Phys.\ Lett.\ B {\bf 738}, 412 (2014)
  [arXiv:1404.5230 [hep-ph]].
  %%CITATION = ARXIV:1404.5230;%%
  %10 citations counted in INSPIRE as of 24 Apr 2015
%\cite{Mohapatra:1979ia}
\bibitem{Mohapatra:1979ia}
  R.~N.~Mohapatra and G.~Senjanovic,
  %``Neutrino Mass and Spontaneous Parity Violation,''
  Phys.\ Rev.\ Lett.\  {\bf 44}, 912 (1980).
  %%CITATION = PRLTA,44,912;%%
  %3731 citations counted in INSPIRE as of 07 Aug 2015
%\cite{Barbieri:1998mq}											
\bibitem{Barbieri:1998mq}
  R.~Barbieri, L.~J.~Hall, D.~Tucker-Smith, A.~Strumia and N.~Weiner,
  %``Oscillations of solar and atmospheric neutrinos,''
  JHEP {\bf 9812}, 017 (1998)
  [hep-ph/9807235].
  %%CITATION = HEP-PH/9807235;%%
  %265 citations counted in INSPIRE as of 07 Aug 2015
%\cite{Agashe:2014kda}
\bibitem{Agashe:2014kda}
  K.~A.~Olive {\it et al.} [Particle Data Group Collaboration],
  %``Review of Particle Physics,''
  Chin.\ Phys.\ C {\bf 38}, 090001 (2014).
  %%CITATION = CHPHD,C38,090001;%%
  %1615 citations counted in INSPIRE as of 07 Aug 2015
%\cite{Babu:2002ex}
\bibitem{Babu:2002ex}
  K.~S.~Babu and R.~N.~Mohapatra,
  %``Predictive schemes for bimaximal neutrino mixings,''
  Phys.\ Lett.\ B {\bf 532}, 77 (2002)
  [hep-ph/0201176].
  %%CITATION = HEP-PH/0201176;%%
  %107 citations counted in INSPIRE as of 07 Aug 2015
%\cite{Karmakar:2015jza}
\bibitem{Karmakar:2015jza} 
  B.~Karmakar and A.~Sil,
  %``Spontaneous CP Violation in Lepton-sector: a common origin for $\theta_{13}$, Dirac CP phase and leptogenesis,''
  arXiv:1509.07090 [hep-ph].
  %%CITATION = ARXIV:1509.07090;%%

%\cite{Kawasaki:2006hm}
\bibitem{Kawasaki:2006hm}
  M.~Kawasaki, F.~Takahashi and T.~T.~Yanagida,
  %``The Gravitino-overproduction problem in inflationary universe,''
  Phys.\ Rev.\ D {\bf 74}, 043519 (2006)
  [hep-ph/0605297].
  %%CITATION = HEP-PH/0605297;%%
  %114 citations counted in INSPIRE as of 09 Aug 2015

%\cite{Kawasaki:2006gs}
\bibitem{Kawasaki:2006gs}
  M.~Kawasaki, F.~Takahashi and T.~T.~Yanagida,
  %``Gravitino overproduction in inflaton decay,''
  Phys.\ Lett.\ B {\bf 638}, 8 (2006)
  [hep-ph/0603265].
  %%CITATION = HEP-PH/0603265;%%
  %134 citations counted in INSPIRE as of 09 Aug 2015
%\cite{Kawasaki:2008qe}
\bibitem{Kawasaki:2008qe}
  M.~Kawasaki, K.~Kohri, T.~Moroi and A.~Yotsuyanagi,
  %``Big-Bang Nucleosynthesis and Gravitino,''
  Phys.\ Rev.\ D {\bf 78}, 065011 (2008)
  [arXiv:0804.3745 [hep-ph]].
  %%CITATION = ARXIV:0804.3745;%%
  %296 citations counted in INSPIRE as of 09 Aug 2015
%\cite{Viel:2005qj}
\bibitem{Viel:2005qj}
  M.~Viel, J.~Lesgourgues, M.~G.~Haehnelt, S.~Matarrese and A.~Riotto,
  %``Constraining warm dark matter candidates including sterile neutrinos and light gravitinos with WMAP and the Lyman-alpha forest,''
  Phys.\ Rev.\ D {\bf 71}, 063534 (2005)
  [astro-ph/0501562].
  %%CITATION = ASTRO-PH/0501562;%%
  %363 citations counted in INSPIRE as of 10 Aug 2015
%\cite{Kohri:2004qu}
\bibitem{Kohri:2004qu} 
  K.~Kohri, M.~Yamaguchi and J.~Yokoyama,
  %``Production and dilution of gravitinos by modulus decay,''
  Phys.\ Rev.\ D {\bf 70}, 043522 (2004)
  [hep-ph/0403043].
  %%CITATION = HEP-PH/0403043;%%
  %36 citations counted in INSPIRE as of 31 Aug 2015

%\cite{Fukushima:2013vxa}
\bibitem{Fukushima:2013vxa}
  H.~Fukushima and R.~Kitano,
  %``Gravitino thermal production revisited and a new cosmological scenario of gauge mediation,''
  JHEP {\bf 1401}, 081 (2014)
  [arXiv:1311.6228 [hep-ph]].
  %%CITATION = ARXIV:1311.6228;%%
  %5 citations counted in INSPIRE as of 09 Aug 2015

%\cite{Buchmuller:1998zf}
\bibitem{Buchmuller:1998zf}
  W.~Buchmuller and T.~Yanagida,
  %``Quark lepton mass hierarchies and the baryon asymmetry,''
  Phys.\ Lett.\ B {\bf 445}, 399 (1999)
  [hep-ph/9810308].
  %%CITATION = HEP-PH/9810308;%%
  %155 citations counted in INSPIRE as of 13 May 2015
%\cite{Jenkins:2008rb}
\bibitem{Jenkins:2008rb}
  E.~E.~Jenkins and A.~V.~Manohar,
  %``Tribimaximal Mixing, Leptogenesis, and theta(13),''
  Phys.\ Lett.\ B {\bf 668}, 210 (2008)
  [arXiv:0807.4176 [hep-ph]].
  %%CITATION = ARXIV:0807.4176;%%
  %47 citations counted in INSPIRE as of 25 août 2015
%\cite{Durnford:2009aba}
\bibitem{Durnford:2009aba}
  C.~Durnford,
  ``Duality and Models of Supersymmetry Breaking.''
  %%CITATION = INSPIRE-1250626;%%



% Please avoid comments such as "For a review'', "For some examples",
% "and references therein" or move them in the text. In general,
% please leave only references in the bibliography and move all
% accessory text in footnotes.

% Also, please have only one work for each \bibitem.


\end{thebibliography}
\end{document}